\documentclass[conference]{IEEEtran}
\IEEEoverridecommandlockouts
\usepackage{amsthm,amsmath,amssymb,amsfonts}
\usepackage[ruled,vlined,linesnumbered,resetcount]{algorithm2e}
\usepackage{array}
\usepackage{balance}  
\usepackage{cite}
\usepackage{hyperref}
\usepackage{cleveref}
\usepackage{comment}
\usepackage{graphicx}
\usepackage{makecell}
\usepackage{mathrsfs}
\usepackage{multirow}
\usepackage{pgfplots}
\usepackage{subfigure}
\usepackage{textcomp}
\usepackage{tikz} 
\usepackage{xcolor}
\usepackage{changepage}
\usepackage{threeparttable}

\usepackage[T1]{fontenc}

\newtheorem{definition}{Definition}
\newtheorem{theorem}{Theorem}
\newtheorem{example}{Example}

\let\oldnl\nl
\newcommand{\nonl}{\renewcommand{\nl}{\let\nl\oldnl}}
\SetKwComment{tcp}{$\triangleright$\ }{}
\SetCommentSty{mycommfont}

\def\BibTeX{{\rm B\kern-.05em{\sc i\kern-.025em b}\kern-.08em
    T\kern-.1667em\lower.7ex\hbox{E}\kern-.125emX}}
    
\begin{document}

\title{Frequency-based Randomization for Guaranteeing Differential Privacy in Spatial Trajectories}

\author{\IEEEauthorblockN{Fengmei Jin$^1$, Wen Hua$^1$, Boyu Ruan$^1$, Xiaofang Zhou$^2$}
	\IEEEauthorblockA{
		{$^1$School of Information Technology and Electrical Engineering, The University of Queensland, Australia}\\
		{$^2$Department of Computer Science and Engineering, The Hong Kong University of Science and Technology, Hong Kong}\\
		{\{fengmei.jin, w.hua, b.ruan\}@uq.edu.au},  {zxf@cse.ust.hk}
	}
}

\maketitle

\begin{abstract}
With the popularity of GPS-enabled devices, a huge amount of trajectory data has been continuously collected and a variety of location-based services have been developed that greatly benefit our daily life. However, the released trajectories also bring severe concern on personal privacy, and several recent studies have demonstrated the existence of personally-identifying information in spatial trajectories. Trajectory anonymization is nontrivial due to the trade-off between privacy protection and utility preservation. Furthermore, recovery attack has not been well studied in the current literature. To tackle these issues, we propose a frequency-based randomization model with a rigorous differential privacy guarantee for trajectory data publishing. In particular, we introduce two randomized mechanisms to perturb the local/global frequency distributions of significantly important locations in trajectories by injecting Laplace noise. We design a hierarchical indexing along with a novel search algorithm to support efficient trajectory modification, ensuring the modified trajectories satisfy the perturbed distributions without compromising privacy guarantee or data utility. Extensive experiments on a real-world trajectory dataset verify the effectiveness of our approaches in resisting individual re-identification and recovery attacks, and meanwhile preserving desirable data utility as well as the feasibility in practice.
\end{abstract}


\vspace{-3mm}
\section{Introduction}
\label{sec:intro}
Nowadays, GPS-enabled devices and location-based applications have become ubiquitous, and an increasing amount of spatiotemporal data has been collected such as vehicle trajectories, phone call records, and user check-ins.
Knowledge discovery from such data promotes the development of many advanced technologies like route planning and recommendation, which bring massive benefits to our daily life. 

However, as a side effect, trajectories become vulnerable and could potentially expose individuals' sensitive information. In particular, a widely-encountered risk in trajectory data is the \textit{linkage} or \textit{re-identification} attack that recognizes an individual from their moving history. Recent studies \cite{de2013unique,chen2017exploiting,jin2019moving,jin2020trajectory} have demonstrated that individuals can be identified with a sufficiently high success rate ($>$ 80\%) by exploring their personalized movement patterns. Hence, to protect individuals from re-identification, many privacy models have been proposed for trajectory data publication including ad-hoc models \cite{liu2012traffic,liu2019dummy} and formal models \cite{abul2010anonymization,gramagliaF15hiding,tu2019protecting,he2015dpt}. Among them, differential privacy (DP) \cite{he2015dpt} is the most powerful mathematical mechanism ensuring that the attacker cannot infer too much about any specific person, which provides a superior theoretical guarantee on data privacy. 
Nevertheless, there are still several notable limitations in existing trajectory DP models: 


\textit{1) Privacy protection vs. utility preservation.} In the context of trajectory data, most approaches achieve protection by adding various types of noise to the entire trajectory, to each point in the trajectory, or to each coordinate of the trajectory point. Unfortunately, excessive modification of trajectories caused by noise injection (no matter at which spatial resolution) dramatically affects the utility of the anonymized data. As a result, trajectories in the differentially private dataset are often useless in practice due to the twisting shapes, unrealistic paths on road networks, sudden changes in direction, etc.~\cite{jiang2013publishing} For example, DPT \cite{he2015dpt}, an innovative DP-based model, uses the synthetic generation framework which captures trajectory movement with prefix trees, adds randomized noise, and then synthesizes trajectories based on the noisy prefix trees. In this way, it offers strong privacy guarantee yet cannot retain the truthfulness of data, since every synthetic trajectory is constructed based on the differentially private movement patterns rather than any real trajectory. 
Therefore, the trade-off between privacy protection and utility preservation has been a major bottleneck for DP-based methods.

\textit{2) Recovery attack.} In fact, it is unnecessary to generate synthetic trajectories from scratch or brutally introduce noise to every single element of the trajectories. Knowing some non-sensitive public locations visited by a user cannot enhance the inference about any personal information. Recently, \cite{de2013unique,jin2019moving} illustrated that only a limited number of spatial points (called \textit{signature} hereafter) contribute the majority to re-identifying an individual from the trajectory dataset. These signature points are the most representative and distinctive in a user's moving history, and hence are sensitive locations (e.g., home and workplace) to be carefully protected. Inspired by this, \cite{jin2020trajectory} attempted to remove all signature points from the original trajectories and preserve the remaining points, which successfully lowered the re-identification risk and meanwhile achieved satisfactory data utility, showing a good balance between privacy protection and utility preservation. However, it is still insufficient to simply discard signature points from the trajectories, since the original traces can be recovered by using the well-developed techniques such as map-matching and path inference. According to our empirical study in \Cref{sec:exp}, 60\% of the original trajectories can be reconstructed from the anonymized dataset, making it still vulnerable to user re-identification. This is called the \textit{recovery attack}, which is also a critical threat to the released trajectory data yet has not attracted enough attention from the research community.



Actually, what makes a location sensitive to individual re-identification is not the location itself but how the user visits that location. In other words, the signature points should be both representative and distinctive in a user's trajectories, i.e., they are frequently visited by a specific user yet rarely by others. Therefore, individual privacy can be successfully protected by distorting the \textit{frequency} of those signature points, where the differential privacy model fits well. In this work, we propose a novel DP model with randomization mechanisms which perturb the frequency distributions of a limited number of signature points from both global and local perspectives, instead of altering the entire trajectories. It naturally protects against re-identification attack, since signatures are the most identifying information in a trajectory. More importantly, it well balances differential privacy guarantee and data utility considering the small scale of signature points, and the perturbed frequency distributions cannot be easily recovered.
Technically, we leverage a non-trivial Laplace mechanism to achieve the differential privacy during global and local frequency perturbation, with a theoretically proved privacy guarantee. In order to modify trajectories based on the distorted frequency distributions and meanwhile minimize the utility loss, we implement several trajectory edit operations, define their utility costs, and formalize trajectory modification as an optimization problem. A greedy solution along with a novel spatial indexing structure are introduced to improve the efficiency of trajectory modification. To sum up, our major contributions in this work are listed below: 




\begin{itemize}
    \item We introduce a novel differential privacy model for trajectories based on frequency perturbation of signatures and non-trivial Laplace mechanism. 
    \item Our DP model not only achieves the balance between privacy protection and utility preservation, but also well resists the recovery attack.
    \item We implement the anonymization procedure to minimize utility loss and design a spatial indexing structure to support efficient trajectory modification.
    \item Extensive experiments on a real-world trajectory dataset verify the superiority of our DP model over existing trajectory protection algorithms. 
\end{itemize}

\Cref{sec:relatedwork} summarizes existing trajectory privacy models; \Cref{sec:dp} introduces our differential privacy model, with a focus on the frequency-based randomization mechanisms; We then explain the technical details of trajectory modification in \Cref{sec:method} and report our experiments in \Cref{sec:exp}; Finally, we conclude this work in \Cref{sec:conclude}.

\section{Related work}\label{sec:relatedwork}
Existing trajectory privacy models can be classified as ad-hoc and formal models \cite{jin2021survey}. Ad-hoc models (e.g., \textit{Mixzone} \cite{liu2012traffic,palanisamy2011mobimix} and \textit{Dummy} \cite{lei2012dummy,kato2012dummy,wu2014novel,liu2019dummy}) are designed specifically based on the characteristics of trajectory data, but fail to provide theoretical privacy guarantee. Formal models extend the privacy guarantee achieved in relational database to trajectory publishing. We elaborate on formal models in this section.



\textit{\textbf{K-anonymity Family:}}
$k$-anonymity-based approaches including $k$-anonymity, $l$-diversity and $t$-closeness aim to make each object indistinguishable within a group of anonymous objects following various principles. As the first work adopting the basic $k$-anonymity to trajectory releasing, \textit{NWA} \cite{abul2008never} and its extended work \textit{W4M} \cite{abul2010anonymization} pursue the ($k$,$\delta$)-anonymity for protecting trajectories where each anonymized trajectory is enforced to co-locate with other $k-1$ trajectories within a cylinder of radius $\delta$. \textit{W4M} adopts the spatiotemporal edit distance to measure the similarity of trajectories and obtain better performance. Instead, \textit{GLOVE} \cite{gramagliaF15hiding} designs another framework by iteratively merging similar trajectory pair with minimum cost in spatiotemporal dimensions until all resulting trips are $k$-anonymous. \textit{KLT} \cite{tu2019protecting} improves over \textit{GLOVE} and protects the semantic information (i.e., the categories of POIs) from exposure by further introducing the $l$-diversity and $t$-closeness, such that each movement in a published trajectory can express various semantic information and is hard to distinguish users. These methods well define the anonymization on trajectory data yet cannot provide desirable privacy guarantee especially when encountering re-identification attack.

\textit{\textbf{Differential Privacy:}}
Differential privacy attempts to make the noisy result derived from a sanitized dataset sufficiently similar to the real answer, such that the adversary cannot obtain too much personal information from the query result while the data is still useful for the purpose of high-utility analysis. Hence, the noise injection should not dramatically blur the original data; otherwise the query result would be too fake to be accepted or the data would become unavailable for use in practice. Existing models achieving differential privacy for trajectories can be classified into following categories.

\subsubsection{Spatial perturbation to original trajectories}
\cite{jiang2013publishing} designs three basic methods for producing noisy trajectory by adding various noises to the entire trajectory, to each sampling point or to every coordinate, respectively. 
To solve the issue of too many crossings appearing in the anonymized trajectories, they further propose the \textit{SDD} mechanism which imposes more requirements for both distance and direction when transiting to the next position. Noises are separately sampled for distance and direction in the angular coordinate system with the help of the \textit{exponential noise} which has also been used in \cite{hua2015differentially,li2017achieving}. Other typical DP models \cite{shao2013publishing,han2018lclean} flexibly employ \textit{randomized response} to inject noises into spatial traces.

\subsubsection{Noisy transition probability for synthetic generation}
Different from the above which injects controlled noises and aims to distort trajectories geographically, \textit{DPT} \cite{he2015dpt} models the trajectories via hierarchical reference systems and encodes the transition information among grid cells at different resolutions with prefix trees. By injecting \textit{Laplace noise} into the prefix trees, the transition probabilities are distorted following differential privacy yet still able to present the movement patterns of the original traces to some extent. Synthetic trajectories are generated from the noisy trees rather than relying on any real trajectory so as to guarantee the differential privacy and preserve original patterns. \textit{SPLT} \cite{vincent2016}, \textit{SafePath} \cite{khalil18}, \textit{DP-STAR} \cite{mehmet19} further extend \textit{DPT} by considering trajectory semantics, temporal information, and various utility features respectively, during noise injection. \textit{AdaTrace} \cite{gursoy2018utility,DBLP:journals/tmc/Gursoy0TY19} fixes the vulnerability of \textit{DPT} by combining differential privacy with attack resilience constraints and a utility-aware synthesizer, resulting in more private and useful trajectories.

As discussed in \Cref{sec:intro}, the major bottleneck of existing differential privacy models is a huge utility loss caused by the excessive changes of trajectories (or even synthetic trips without keeping the record-level truthfulness). We aim to balance privacy protection and data utility in this work. 

\section{Differential Privacy Model}
\label{sec:dp}
In this section, we first introduce some basic concepts of the formal differential privacy and the Laplace mechanism. We then detail our privacy model which provides a desirable differential privacy guarantee by adding non-trivial Laplace noise to the local and global frequency distributions of trajectories, respectively. It is worth noting that these two perturbation mechanisms are independent and can be applied either individually or collectively, supported by the composition property of differential privacy (as introduced in \Cref{dp:composition}). \Cref{tab:notation} summarizes some major notations used in this work.

\vspace{-2mm}
\begin{table}[hbt]
    \centering
    \renewcommand{\arraystretch}{1.1}
    \caption{Summary of notations}
    \label{tab:notation}
    \begin{tabular}{l|p{4.39cm}}
        \hline
        \textbf{Notation} & \textbf{Definition} \\ \hline
        $\mathcal{M}$ & an $\epsilon$-differentially private mechanism ($\epsilon$ is the total privacy budget) \\ \hline
        $D=\{\tau_1,\dots,\tau_{|D|}\}$ & a dataset containing $|D|$ trajectories generated by $|D|$ moving objects \\ \hline
        $D^*=\mathcal{M}(D)$ & the randomized trajectory dataset \\\hline
        $\mathcal{S}_m=\{s_m(\tau_1),\dots,s_m(\tau_{|D|})\}$ & top-$m$ signatures for each trajectory in $D$ ($m$ is a user-defined parameter) \\ \hline
        $\mathcal{P}=\{p_1,p_2,\dots,p_d\}$ & the set of all distinct points composing all signatures in $\mathcal{S}_m$ ($d\leq |D|\times m$) \\\hline
        $\mathcal{F}(\tau)=\langle f_1,f_2,\dots,f_{|\tau|}\rangle$ & the PF distribution over trajectory $\tau$. Each point $p_i$ has its frequency $f_i$ \\\hline
        $\mathcal{F}^*(\tau)=\langle f_1^*,f_2^*,\dots,f_{|\tau|}^*\rangle$ & the perturbed PF distribution of $\tau$ \\\hline
        $\mathcal{L}=\langle l_1,l_2,\dots,l_d\rangle$ & the TF distribution over $\mathcal{P}$ where $l_i$ denotes the number of trajectories in $D$ passing through point $p_i$ \\\hline
        $\mathcal{L}^*=\langle l_1^*,l_2^*,\dots,l_d^*\rangle$ & the perturbed TF distribution over $\mathcal{P}$ \\\hline
    \end{tabular}
    \vspace{-2mm}
\end{table}



\subsection{Differential Privacy}\label{subsec:dp}
\begin{definition}[$\epsilon$-differential privacy]\label{def:dp}
    A randomized mechanism $\mathcal{M}$ provides $\epsilon$-differential privacy if for any two adjacent databases $D$ and $D^\prime$ differing in at most one record, and for every possible output $S\in Range(\mathcal{M})$,
    \begin{equation}
        \Pr[\mathcal{M}(D)\in S] \leq e^\epsilon \cdot \Pr[\mathcal{M}(D^\prime)\in S]
    \end{equation}
    where $\epsilon$ is called as the privacy budget.
\end{definition}

\begin{definition}[Sensitivity]
    Given a query function $\phi$, the sensitivity of $\phi$ is defined as the maximum of the difference in the output value of $\phi$ if $D$ and $D^\prime$ differ in at most one record, i.e.,
    \begin{equation}
        \Delta \phi = \max_{D,D^\prime} \|\phi(D)-\phi(D^\prime)\|_1
    \end{equation}

\end{definition}

In \cite{dwork2006calibrating}, the Laplace mechanism was first proposed to achieve $\epsilon$-differential privacy by injecting noise drawn from the Laplace distribution into the query outputs, where the scale of the adopted Laplace distribution $\lambda$ is determined by the sensitivity of the query function $\Delta \phi$ and the privacy budget $\epsilon$:
\begin{definition}[Laplace mechanism]
    Given a query function $\phi$ and the privacy budget $\epsilon$, a randomized mechanism $\mathcal{M}$ provides the $\epsilon$-differential privacy by sampling i.i.d variables from the Laplace distribution $Lap(\frac{\Delta \phi}{\epsilon})$$\footnote{In general, a Laplace distribution is denoted by $Lap(\mu,\lambda)$ where $\mu$ is the mean and $\lambda$ is the scale. A simplified version is $Lap(\lambda)$ representing the distribution is centered by 0 with the scale $\lambda$, as the $\mu$ is omitted.}$ and adding them to the query answers.
\end{definition}

An important property of differentially private mechanisms is the ``\textit{sequential composition theorem}'' \cite{dwork2014algorithmic}, which lays the foundation for sequentially combining multiple mechanisms without compromising the eventual privacy guarantee:
\begin{theorem}[Sequential composition theorem]\label{dp:composition}
    Assume there are multiple randomized mechanisms $\mathcal{M}_i$, each of which can achieve $\epsilon_i$-differential privacy. Thus, the 
    sequential combination of these mechanisms will become an $\epsilon$-differentially private model with the privacy budget $\epsilon = \sum_i \epsilon_i$. 
\end{theorem}

\vspace{-2mm}
\subsection{Frequency-based Randomization Mechanism}
\begin{definition}[Trajectory]\label{def:traj}
    A trajectory is denoted by a sequence of spatial points ordered chronologically, i.e., $\tau=\{p_1,\dots,p_{|\tau|}\}$, 
    and each object is associated with a single trajectory representing its entire moving history. 
\end{definition}
Dataset $D$ contains $|D|$ trajectories generated by $|D|$ moving objects, denoted as $D=\{\tau_1, \tau_2, ..., \tau_{|D|}\}$. We say $D$ and $D^\prime$ are adjacent datasets only if they differ in at most one trajectory. 
Our frequency-based model achieves $\epsilon$-differential privacy using the Laplace mechanism. In the following, we first explain some fundamental concepts and then introduce the randomization mechanisms for perturbing global and local frequency distributions of trajectories, respectively.

\subsubsection{\textbf{Trajectory Signature and Frequency Distributions}}
Existing mechanisms achieving differential privacy for trajectory data usually inject a large amount of noises into each trajectory or the entire dataset with various approaches (e.g., \cite{jiang2013publishing,he2015dpt}), making the anonymized dataset useless in practice with a huge utility damage. 
Inspired by \cite{jin2019moving}, we observe that \textit{signature} points carry the majority of identifying information in a user's trajectories, indicating that it suffices to anonymize only the signature points in order to protect trajectories against re-identification attack. Hence, our DP model is designed based on the selected signature points, which greatly preserves the utility of the remaining points while providing the same differential privacy guarantee.

\begin{figure}[htb]
    \centering
    \includegraphics[width=0.42\textwidth]{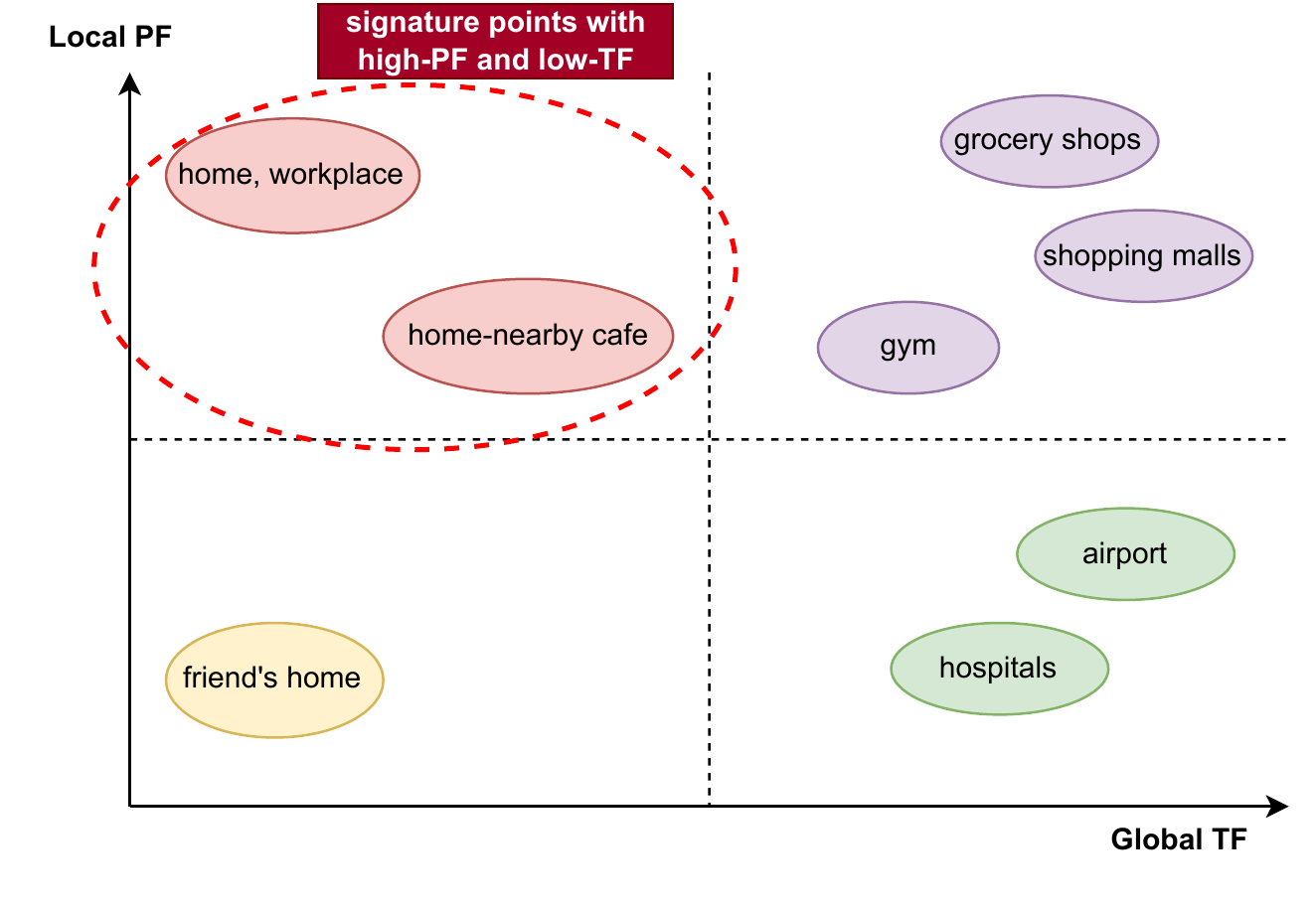}
    \vspace{-5mm}
    \caption{An illustration of signature points with high PF and low TF.}
    \label{fig:sigpoints}
\end{figure}

Ideally, signatures should be both representative and distinctive in a user's trajectories, i.e., they should be frequently visited by a specific user but others rarely go there (as exemplified in \Cref{fig:sigpoints}). Accordingly, our privacy model consists of two independent components, each of which achieves differential privacy for the trajectory dataset by blurring the representativeness and distinctiveness of signature points, respectively. In particular, we utilize two types of frequencies to identify signature points for each trajectory in dataset $D$:
\begin{itemize}
    \item \textit{Point Frequency (PF)}: The signature point should be ubiquitous (representative) in a user's trajectory. We define PF $f_p$ as the total number of times a point $p$ occurring in a trajectory $\tau$, and the representativeness of $p$ in $\tau$ is measured by $\frac{f_p}{|\tau|}$ where $|\tau|$ is the total number of points in $\tau$. Obviously, the higher PF the point has, the more representative it is with regard to a user;
    \item \textit{Trajectory Frequency (TF)}: The signature point should be unique (distinctive) for a specific user. We define TF $l_p$ of a point $p$ as the number of trajectories in $D$ passing through $p$ at least once, and the distinctiveness of $p$ in a dataset $D$ is computed as $\log(\frac{|D|}{l_p})$ where $|D|$ is the total number of trajectories (objects) in $D$. Hence, the lower TF the point has, the more distinctive it is within $D$.
\end{itemize}

\vspace{-2mm}

Each point $p$ in trajectory $\tau$ is weighted as the product of its representativeness and distinctiveness. We extract the top-$m$ points with the largest weights as the signatures of every trajectory, denoted as $\mathcal{S}_m=\{s_m(\tau_1),s_m(\tau_2),\dots,s_m(\tau_{|D|})\}$. All points occurring in at least one signature are grouped into a candidate point set $\mathcal{P}=\{p_1,p_2,\dots,p_d\}$, where $d$ represents the dimensionality of these distinct signature points. In our differential privacy model, we only distort the selected signature points by injecting novel Laplace noises to the local PF distribution of every single trajectory $\mathcal{F}(\tau)=\langle f_1,f_2,\dots,f_{|\tau|}\rangle$ with the privacy budget $\epsilon_L$, and to the global TF distribution $\mathcal{L}=\langle l_1,l_2,\dots,l_d\rangle$ with the privacy budget $\epsilon_G$, rather than altering the entire trajectory data in geographical space.
Based on the ``sequential composition theorem'' in \Cref{dp:composition}, $\epsilon$-differential privacy can be achieved by combining both local and global perturbations with $\epsilon=\epsilon_L+\epsilon_G$. 
After perturbing the local/global distributions, we then modify the trajectories according to the new distributions aiming to minimize utility loss, which will be discussed in \Cref{sec:method}.

\subsubsection{\textbf{Global TF Randomization Mechanism}}
Global randomization applies to the Trajectory Frequency (TF) distribution of all top-$m$ signature points, i.e., $\mathcal{P}=\{p_1,p_2,\dots,p_d\}$. Intuitively, a smaller TF value implies that this location is quite unique to some specific users since fewer individuals have visited there, while a point with a larger TF is more likely to be a hotspot (e.g., shopping mall, airport, train station, etc.) embedding less personal information. Hence, globally perturbing the TF distribution of signature points is important for blurring those super distinctive locations and preventing the leakage of personal features when publishing the trajectories.

\begin{algorithm}[hbt]
    \small
    \caption{Global TF Randomization Mechanism}
    \label{algo:global}
    \KwIn{$D$: A dataset with $|D|$ trajectories; $m$: the signature size; $\mathcal{P}$: the set of top-$m$ signature points; $\epsilon_G$: the privacy budget for global perturbation}
    \KwOut{$D_G^*$: The TF-perturbed dataset}
    $\mathcal{L}\leftarrow$ Build-TF-Distribution $(D,\mathcal{P},m)$\tcp*[r]{over $\mathcal{P}$}
    \For{$\forall l_j\in\mathcal{L}$}{
        $\eta \thicksim Lap(\frac{1}{\epsilon_G})$;\\
        $l_j^*\leftarrow l_j+\eta$;\\
        $l_j^*\leftarrow$ Round$(l_j^*, [0,|D|]$) \tcp*[r]{post-processing: round the noisy TF value to a proper integer range}
    }
    $\mathcal{L}^*\leftarrow\langle l_1^*,l_2^*,\dots,l_d^*\rangle$;\\
    $D_G^*\leftarrow$ GlobalEdit$(D,\mathcal{L}^*)$;\\
    \Return{$D_G^*$};
\end{algorithm}

As shown in \Cref{algo:global}, we generate a global TF list $\mathcal{L}$ over the set of top-$m$ signature points $\mathcal{P}$, where each element $l_j$ represents the TF value for point $p_j$ (line 1). Consider a point counting query $\phi(D,p_j)$ which is exactly asking for the TF value of the given point $p_j$. The difference of the query answers on two adjacent databases $D$ and $D^\prime$ differing in one single trajectory is at most $\pm 1$ due to the existence or non-existence of that trajectory, resulting in the sensitivity as $\Delta \phi=1$. Therefore, we follow the Laplace mechanism introduced in \Cref{subsec:dp} to obtain $\epsilon_G$-DP guarantee by directly adding noise drawn from $Lap(1/\epsilon_G)$ to each element $l_j$ (lines 3-5). Note that neither TF nor PF of a point can be a negative or float value due to their semantics. Thus, we always round the noisy values to zero or an integer range if needed (line 5). Such a post-processing operation will not damage the guaranteed differential privacy \cite{dwork2014algorithmic}. After noise injection, we conduct the global alteration to modify trajectories based on the perturbed TF distribution $\mathcal{L}^*$ (line 7, detailed in \Cref{sec:method}).

\subsubsection{\textbf{Local PF Randomization Mechanism}}
The purpose of perturbing global TF distribution is to distort the distinctiveness of top-ranked signature points in a global view. On the other hand, some locations repeatedly appearing in a user's trajectory can also expose the personal identity. To this end, our local mechanism aims to perturb the Point Frequency (PF) distribution for each trajectory, in particular, reducing the occurrence of personally-identifying locations to dilute their representativeness and increasing the occurrence of other insensitive points to introduce randomness in a reasonable way.

However, the property of the commonly-used Laplace mechanism $Lap(\mu=0,\lambda=\frac{\Delta \phi}{\epsilon})$ determines that the probability of sampling positive noise is exact equal to that of negative noise since the distribution center is zero. To achieve the goal of reducing or increasing PF for some specific points with a higher probability, we design a novel local randomization mechanism by using the Laplace distribution with non-zero mean, i.e., $Lap(\mu\neq 0,\lambda=\frac{\Delta \phi}{\epsilon})$. We will theoretically prove in \Cref{theorem:local} that using such kind of non-trivial Laplace mechanism still preserves the $\epsilon$-differential privacy. 

\vspace{-2mm}
\begin{algorithm}[ht]
    \small
    \caption{Local PF Randomization Mechanism}
    \label{algo:local}
    \KwIn{$D$: A dataset with $|D|$ trajectories; $\mathcal{F}$: the original PF distribution of each trajectory; $m$: the number of signature points to be perturbed; $PL$: the selected point list for each trajectory; $\epsilon_L$: the privacy budget for local perturbation}
    \KwOut{$D_L^*$: The PF-perturbed dataset}
    \For{$\forall \tau_i \in D$}{
        \tcc{Stage-1: perturb PF for top-$m$ ranked points}
        $\bar{\mu}\leftarrow 0$;\\
        \For{$k\in [1,m]$}{
            $p_k\leftarrow$ the $k$-th point in $PL(\tau_i)$;\\
            $f_k \leftarrow \mathcal{F}(\tau_i, p_k)$\tcp*[r]{the original PF}
            $\eta \thicksim Lap(-f_k,\frac{1}{\epsilon_L})$; \\
            $f_k^*\leftarrow f_k + \eta$\tcp*[r]{noise injection}
            $f_k^*\leftarrow$ RoundInt$(f_k^*)$\tcp*[r]{round to an integer}
            $f_k^*\leftarrow \max(f_k^*,0)$\tcp*[r]{round negative to zero}
            $\bar{\mu} \leftarrow \bar{\mu} + (f_k^* - f_k)$\tcp*[r]{sum up the actual noise}
        }
        $\bar{\mu}\leftarrow\bar{\mu}/m$\tcp*[r]{the avg. truly added noises in Stage-1}
        \tcc{Stage-2: perturb PF for remaining $m$ points}
        \For{$k\in [m+1,2m]$}{
            $p_k\leftarrow$ the $k$-th point in $PL(\tau_i)$;
            $f_k \leftarrow \mathcal{F}(\tau_i,p_k)$;\\
            $\eta \thicksim Lap(-\bar{\mu},\frac{1}{\epsilon_L})$; 
            $f_k^*\leftarrow f_k + \eta$;\\
            $f_k^*\leftarrow$ RoundInt$(f_k^*)$;
            $f_k^*\leftarrow \max(f_k^*,0)$; \\
        }
        $\mathcal{F}^*(\tau_i)\leftarrow\langle f_1^*, f_2^*\dots f_{2m}^*\rangle$;\\
    }
    $D_L^*\leftarrow$ LocalEdit$(D,\mathcal{F}^*)$;\\

    \Return{$D_L^*$};
\end{algorithm}

\vspace{-2mm}
\paragraph{Algorithm Description} The local perturbation covers two stages of noise injection, which can not only blur the personally sensitive information in a user's moving history but also ensure the total injected noises will not greatly affect the trajectory length. 
Initially, each trajectory is assigned to a list with $2m$ selected points where $m$ denotes the signature size. These points are selected by sequentially picking from the intersection set of its top-ranked signature and the set $\mathcal{P}$ and randomly sampling from its remaining points until the size reaches $2m$.
We then probabilistically decrease the frequency of top-$m$ points in Stage-1 for diluting their representativeness, and meanwhile increase the occurrence of other $m$ points in Stage-2 for keeping trajectory cardinality to a large extent.

As presented in \Cref{algo:local}, for each point $p_k$ in Stage-1, we inject the noise $\eta \thicksim Lap(-f_k,\frac{1}{\epsilon_L})$ into its original frequency value $f_k$ (lines 4-7). The reason for using Laplace distribution with negative mean (i.e., $-f_k$) is to sample negative noises with a higher probability so as to reduce the occurrence of top-$m$ signature points to the greatest extent. 
Regarding the scale $\lambda$, a point-counting query $\phi(p,\tau)$, which estimates the total number of occurrence in $\tau$ for the point $p$, would naturally result in the sensitivity $\Delta \phi=1$. We use a post-processing operation to round the noisy frequency to the closest integer (line 8) or to zero if it is negative (line 9) \cite{dwork2014algorithmic}.
Similar operations are conducted on the remaining $m$ points (lines 12-17). The main difference is the Laplace distribution used in Stage-2 is $-\bar{\mu}$, which is the average of the truly added noises in Stage-1. Regarding all the other points that are not as important as signature points in terms of reidentification ability, they remain unchanged to preserve more data utility. Finally, we make the trajectories satisfying the perturbed PF distribution $\mathcal{F}^*$ using the intra-trajectory modification (line 20), which will be detailed in \Cref{sec:method}.

\paragraph*{\underline{The Importance of Stage-2}}
The most important goal of the local randomization mechanism is to reduce the occurrence of top-ranked signature points, as what we expect to achieve in Stage-1. 
However, purely conducting Stage-1 without the help of Stage-2 
would dramatically influence the cardinality of the resulting trajectory. More specifically, it would result in a huge drop in the total number of points, leading to poor utility of the anonymized dataset.
Our mechanism avoids this issue with the help of Stage-2 modification, which induces both raise and drop of trajectory length in a relatively even manner. Again, our prior choice for the frequency-increasing task is other top-ranked signature points appearing in $\mathcal{P}$ as well. Compared to other insignificant points, it is more convincing to increase the occurrence of these points considering their PF and TF weights. Plus, if the selected points appear in $\mathcal{P}$ implying its significance to other objects, then raising its frequency would bring confusing message as additional benefits.

\paragraph{Privacy Analysis} We first prove that a generalized Laplace mechanism using Laplace distribution $Lap(\mu,\lambda)$ with a non-zero mean $\mu$ and the scale $\lambda=\frac{\Delta \phi}{\epsilon}$ can still guarantee the $\epsilon$-differential privacy, same as that of the commonly-used Laplace distribution centered by zero. Then, we leverage it to illustrate that our local mechanism can strictly provide the $\epsilon_L$-differential privacy guarantee to the trajectory data. 
\begin{theorem}\label{theorem:Laplace}
    A randomized mechanism $\mathcal{M}$ sampling noises from the Laplace distribution $Lap(\mu\neq 0,\lambda=\frac{\Delta \phi}{\epsilon})$ still preserves $\epsilon$-differential privacy.
\end{theorem}
\begin{proof}
    Recall that the adopted Laplace distribution has a probability density function: $Lap(x|\mu,\lambda)=\frac{1}{2\lambda}\exp(\frac{-|x-\mu|}{\lambda})$.
    Assume $\boldsymbol{x}=(x_1,x_2,\cdots,x_n)$ and $\boldsymbol{y}=(y_1,y_2,\cdots,y_n)$ are two adjacent input variables differing at most one dimension, such that $\|x-y\|\leq 1$. Let $\phi(\cdot)$ be some query functions $\phi: \mathbb{N}^n\rightarrow \mathbb{R}^m$. 
    We compare the probability of outputting the same arbitrary result $\boldsymbol{z}\in\mathbb{R}^m$ for them:
    \vspace{-1mm}
    \begin{equation*}
        \small
        \begin{split}
            &\frac{\Pr(\mathcal{M}(\boldsymbol{x})=\boldsymbol{z})}%
            {\Pr(\mathcal{M}(\boldsymbol{y})=\boldsymbol{z})} = %
            \prod_{i=1}^m\left(\frac{\exp(-\frac{\epsilon\big||z_i-\phi(x)_i|-\mu\big|}{\Delta \phi})}%
                {\exp(-\frac{\epsilon\big||z_i-\phi(y)_i\mid-\mu\big|}{\Delta \phi})}\right) \\
            &\ = \prod_{i=1}^m\exp\left(\frac{\epsilon(\big||z_i-\phi(y)_i|-\mu\big|-\big||z_i-\phi(x)_i|-\mu\big|)}{\Delta \phi}\right) \\
            &\ \leq \prod_{i=1}^m\exp\left(\frac{\epsilon\big||z_i-\phi(y)_i|-|z_i-\phi(x)_i|\big|}{\Delta \phi}\right) \\
            &\ \leq \prod_{i=1}^m\exp\left(\frac{\epsilon|\phi(x)_i-\phi(y)_i|}{\Delta \phi}\right) \\
            &\ =\exp\left(\frac{\epsilon\|\phi(\boldsymbol{x})-\phi(\boldsymbol{y})\|_1}{\Delta \phi}\right) \leq \exp(\epsilon)
        \end{split}
    \end{equation*}
\end{proof}

\begin{theorem}\label{theorem:local}
    The local perturbation mechanism described in \Cref{algo:local} provides $\epsilon_L$-differential privacy to a trajectory $\tau$.
\end{theorem}

\begin{proof}
    Let $\tau$ and $\tau^\prime$ be adjacent trajectories differing at most one point.
    Let $\phi(\cdot)$ be a point-counting query function, which returns the frequency of $p_k$ in $\tau$ as the query answer, denoted as $\phi(\tau)_k$ for short. Apparently, $\|\phi(\tau)-\phi(\tau^\prime)\|_1\leq 1$.
    According to the local perturbation mechanism, each point $p_k$ belonging to the top-$m$ signature of a specific trajectory is distorted in Stage-1 by injecting noises drawn from $Lap(-f_k,\frac{1}{\epsilon_L})$ into its original frequency $f_k$, while the noises sampled for the remaining $m$ points are from $Lap(-\bar{\mu},\frac{1}{\epsilon_L})$ where $\bar{\mu}$ is the average noises truly added in previous stage. In another word, two Laplace distributions share the same scale $1/\epsilon_L$ but have various mean values. Based on \Cref{theorem:Laplace}, we claim that:
    
    \begin{equation*}
        \small
        \begin{split}
            &\frac{\Pr(\mathcal{M}(\tau)=\tilde{\tau})}{\Pr(\mathcal{M}(\tau^\prime)=\tilde{\tau})} = \prod_{k=1}^{m}\left(\frac{\exp(-\epsilon_L\big||\tilde{\tau}_k-\phi(\tau)_k|-f_k\big|)}{\exp(-\epsilon_L\big||\tilde{\tau}_k-\phi(\tau^\prime)_k|-f_k\big|)}\right) \\
            &\qquad \times\prod_{k=m+1}^{2m}\left(\frac{\exp(-\epsilon_L\big||\tilde{\tau}_k-\phi(\tau)_k|+\bar{\mu}\big|)}{\exp(-\epsilon_L\big||\tilde{\tau}_k-\phi(\tau^\prime)_k|+\bar{\mu}\big|)}\right) \\
            &\ \leq \prod_{k=1}^m\exp\left(\epsilon_L\big||\tilde{\tau}_k-\phi(\tau^\prime)_k|-|\tilde{\tau}_k-\phi(\tau)_k|\big|\right) \\
            &\qquad \times\prod_{k=m+1}^{2m}\exp\left(\epsilon_L(|\tilde{\tau}_k-\phi(\tau^\prime)_k|-|\tilde{\tau}_k-\phi(\tau)_k|)\right) \\
            &\ \leq \prod_{k=1}^{2m}\exp\left(\epsilon_L|\phi(\tau)_k-\phi(\tau^\prime)_k|\right) \\
            &\ = \exp(\epsilon_L\|\phi(\tau)-\phi(\tau^\prime)\|_1) \leq \exp(\epsilon_L)
        \end{split}
    \end{equation*}
\end{proof}

\section{Trajectory Modification}\label{sec:method}
After we obtain the globally-perturbed trajectory frequency (TF) distribution $\mathcal{L}^*$ and locally-perturbed point frequency (PF) distribution $\mathcal{F}^*$, our next task is to modify the original trajectories based on the following requirements:
\begin{itemize}
    \item \textit{Adapt to the perturbed distributions}: 1) to enforce each trajectory satisfy its perturbed local PF distribution; 2) to enforce the modified trajectory dataset follow the perturbed global TF distribution. The guaranteed differential privacy will not be compromised only when these two frequency distributions are sufficiently satisfied.
    \item \textit{Minimize the utility loss}: Under the premise of satisfying the above requirement, the fewer changes we make to the trajectories, the less utility we lose after anonymization. Hence, our goal of the modification is to minimize the utility loss caused by various trajectory edit operations.
\end{itemize}

Hence, we design the intra-trajectory modification for the local mechanism to modify each trajectory until it satisfies its perturbed PF distribution, and the inter-trajectory modification on multiple trajectories to make the whole dataset meets the perturbed global TF distribution. We will first introduce various trajectory edit operations with the utility loss, define the optimal trajectory modification problems, and finally describe our algorithms to support efficient trajectory modification.

\subsection{Trajectory Edit Operations and Utility Loss}
\begin{figure}[hbt]
    \centering
    \includegraphics[width=0.48\textwidth]{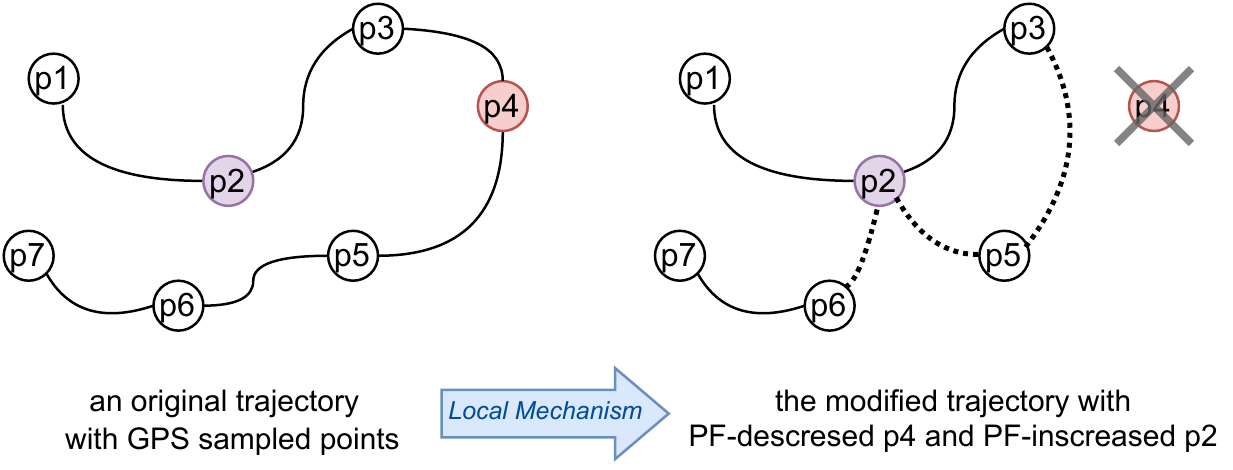}
    \caption{An example of trajectory editing.}
    \label{fig:eg-local}
\end{figure}

The change of point $q$'s frequency (either global TF or local PF) only includes two cases: 1) frequency increasing asks inserting more $q$ into a trajectory (or more); 2) frequency decreasing requires deleting the existing $q$ from a trajectory (or more). Here, the injected noise determines how many times point $q$ should be inserted or deleted. Thus, we introduce two basic operations for trajectory modification: $\mathcal{OP}_i$ and $\mathcal{OP}_d$ denoting the insertion and deletion, respectively. \Cref{fig:eg-local} illustrates an example where we delete point $p_4$ and insert a new occurrence of point $p_2$ (to its closest trajectory segment $\langle p_5, p_6 \rangle$). The trajectory will be modified from $\tau=\{p_1,p_2,p_3,p_4,p_5,p_6,p_7\}$ to $\tau^*=\{p_1,p_2,p_3,p_5,p_2,p_6,p_7\}$.

Another critical issue is how to compute the utility loss caused by performing these edit operations. Recall that our goal is privacy-preserving trajectory publishing, which requires the resulting dataset should be able to serve any generic query/mining task with the controlled amount of information leakage. Thus, our utility loss is defined in a basic distance-based manner rather than a task-specific way. 
In particular, the distance between a point $q$ and a line segment $s=\langle p_x, p_y\rangle$ is computed as the minimum distance between $q$ and the closest point $\bar{p}$ on $s$:
\begin{equation}
    \label{eq:segd}
    dist(q,s) = \min_{\bar{p}\in s} dist(q, \bar{p})
\end{equation}


\begin{definition}[Utility Loss of Point Insertion]
    Inserting a point $q$ into a line segment $s$ will cause a utility loss of: $\mathbb{L}[\mathcal{OP}_i(q,s)] = dist(q,s)$, where $s=\langle p_x,p_y\rangle$ is an arbitrary line segment composed of two endpoints $p_x,p_y$ in $\tau$. 
\end{definition}

Hence, the cost of inserting $q$ into a trajectory $\tau$ once with minimal utility loss is defined as $\mathbb{L}[\mathcal{OP}_i(q,\tau)]=\min_{s\in\tau}\mathbb{L}[\mathcal{OP}_i(q,s)]$. 
Naturally, if the insertion happens $\Delta$ times, the utility loss will be accumulated: $\mathbb{L}[\mathcal{OP}_i(q,\tau,\Delta)]=\sum_{i=1}^{\Delta} \mathbb{L}[\mathcal{OP}_i(q,s_i)]$, where $s_i$ is the trajectory segment selected from $\tau$ for the $i$-th insertion. 

\begin{definition}[Utility Loss of Point Deletion]
    The utility loss of deleting a point $q$ from an existing trajectory segment $s=\langle p_x, q, p_y\rangle$ in $\tau$ is calculated as: $\mathbb{L}[\mathcal{OP}_d(q,s)]=dist(q,s^\prime)$, where $s^\prime=\langle p_x,p_y\rangle$ is the line segment by removing $q$ and reconnecting two endpoints $p_x, p_y$ of $s$. 
\end{definition}

Similarly, the cost of deleting $q$ from a trajectory $\tau$ $\Delta$ times is accumulated as: $\mathbb{L}[\mathcal{OP}_d(q,\tau,\Delta)]=\sum_{i=1}^{\Delta} \mathbb{L}[\mathcal{OP}_d(q,s_i)]$, where $s_i$ is the trajectory segment selected from $\tau$ for the $i$-th deletion. If a point is forced to disappear from a trajectory, we simply remove all its occurrences with the overall utility loss as: $\mathbb{L}[\mathcal{OP}_d(q,\tau)]=\sum_{s\in\tau} \mathbb{L}[\mathcal{OP}_d(q,s)]$.

\subsection{Optimal Trajectory Modification}
Based on the above edit operations, we formally define the problems of the inter- and intra-trajectory modification, respectively. Then we explain how to solve them by reducing to a $K$-nearest trajectory (segment) search problem. 

\vspace{1mm}
\subsubsection{\textbf{Inter-trajectory Modification in Global Mechanism}}
\begin{definition}[Inter-trajectory Modification]
    Given a set of trajectories $D$ and the original/perturbed TF distribution over the selected point set $\mathcal{P}$, two types of points with noisy TF should be processed during inter-trajectory modification: 1) insert a point $q$ into $\Delta l_q$ selected trajectories \textbf{at least once} to make its TF increasing $\Delta l_q$ times; and 2) \textbf{completely} delete a point $q$ from $\Delta l_q$ trajectories to make its TF decreasing $\Delta l_q$ times. 
    Here, $\Delta l_q=|l_q^*-l_q|$ represents the growth/drop of $q$'s TF value after global perturbation. Meanwhile, the minimum utility loss of the dataset is expected when altering trajectories.
\end{definition}
The core of this task is to select the most appropriate trajectories for the insertion/deletion, such that the utility loss caused by modifying them can be minimized. That is, we can solve it as a \textbf{$K$-nearest trajectory search} problem where the ``nearest'' can be defined as below:
\begin{definition}[$K$-Nearest Trajectory Search]
    Given a point $q$ and a trajectory set $D$, the TF-increasing task, i.e., inserting $q$ into $\Delta l_q$ selected trajectories of $D$ at least once, aims to find a list of top-$\Delta l_q$ trajectories with the minimum insertion utility loss for $q$, where $\Delta l_q=|l_q^*-l_q|$, such that:
    \begin{align*}
        & \mathbb{L}[\mathcal{OP}_i(q,\tau)] \leq \mathbb{L}[\mathcal{OP}_i(q,\tau^\prime)] \\
        & \forall \tau\in NT_q^+, \forall \tau^\prime\in \{D\setminus NT_q^+\}
    \end{align*} 
    where $NT_q^+$ contains the $\Delta l_q$-nearest trajectory selected from $D$ for the increasing TF of $q$.
    \vspace{1mm}

    By contrast, the TF-decreasing task requires completely deleting $q$ from $\Delta l_q$ trajectories in $D$, which is equivalent to finding a list of top-$\Delta l_q$ trajectories with the minimum complete deletion loss for $q$, such that:
    \begin{align*}
        & \mathbb{L}[\mathcal{OP}_d(q,\tau)] \leq \mathbb{L}[\mathcal{OP}_d(q,\tau^\prime)]\\ 
        & \forall \tau\in NT_q^-,~\forall \tau^\prime\in \{D\setminus NT_q^-\}
    \end{align*} 
    where $NT_q^-$ records the selected trajectories for point deletion with the size of $|NT_q^+|=\Delta l_q$.
\end{definition}

\subsubsection{\textbf{Intra-trajectory Modification in Local Mechanism}}
\begin{definition}[Intra-trajectory Modification]
    Given a trajectory $\tau$ and its original/perturbed PF distribution, denoted by $\mathcal{F}(\tau)$ and $\mathcal{F}^*(\tau)$ respectively, the goal of intra-trajectory modification is to insert (resp. delete) a point $q$ into (resp. from) $\tau$ for $\Delta f_q$ times, where $q$ denotes every point whose frequency in $\tau$ increases (resp. decreases) after local perturbation and $\Delta f_q=|f^*_q-f_q|$, while, more importantly, minimizing the utility loss of $\tau$ caused by the modification. 
\end{definition}

Again, considering our optimization goal, when editing a point $q$ to satisfy its perturbed frequency, it is reasonable and practical to pick the top-ranked nearest segments with respect to $q$, which enables us to reduce the intra-trajectory modification to a \textbf{$K$-nearest segment search} problem:
\begin{definition}[$K$-Nearest Segment Search]
    Given a point $q$ and a trajectory $\tau$, the task of inserting $q$ into $\tau$ for $\Delta f_q$ times can be regarded as finding a list of top-$\Delta f_q$ nearest trajectory segments with the minimum insertion utility loss for $q$, where $\Delta f_q=|f^*_q-f_q|$, such that:
    \begin{align*}
        & \mathbb{L}[\mathcal{OP}_i(q,s)] \leq \mathbb{L}[\mathcal{OP}_i(q,s^\prime)] \\
        & \forall s\in NS_q^+~,~\forall s^\prime\in \{S(\tau)\setminus NS_q^+\}
    \end{align*} 
    where $NS_q^+$ contains the $\Delta f_q$-nearest trajectory segments to $q$, and $S(\tau)$ represents the set of all possible trajectory segments composed of any two consecutive points in $\tau$.
    \vspace{1mm}

    Similarly, deleting $q$ from $\tau$ for $\Delta f_q$ times is equivalent to finding a list of top-$\Delta f_q$ trajectory segments passing $q$, namely, $NS_q^-=\{s=\langle p_x, q, p_y\rangle | s\in\tau\}$, such that:
    \begin{align*}
        & \mathbb{L}[\mathcal{OP}_d(q,s)] \leq \mathbb{L}[\mathcal{OP}_d(q,s^\prime)] \\
        & \forall s\in NS_q^-,~\forall s^\prime\in \{S(\tau,q)\setminus NS_q^-\}
    \end{align*} 
    where $NS_q^-$ records the selected trajectory segments with the size of $|NS_q^+|=\Delta f_q$, and $S(\tau,q)$ represents the set of all trajectory segments composed of any three consecutive points in $\tau$ and the middle one is the target point $q$.
\end{definition}

\subsection{Efficient Hierarchical Grid Index}
\label{subsec:index}
The most straightforward solution for the above $K$-nearest trajectory segment search problem is linearly scanning the whole dataset for every frequency-perturbed point, which is quite time-consuming for both modification tasks since the computation complexity of such naive linear scan reaches $O(dn\bar{l})$ and $O(mn\bar{l})$ for the inter- and intra-trajectory modification respectively. Here, $d$ denotes the total dimensionality of $\mathcal{P}$ (reaching $O(mn)$ in the worst case); $n$ and $\bar{l}$ are the dataset cardinality and average trajectory length respectively. 

Considering the locality property of geographic data, adopting a spatial index would potentially speed up the nearest neighbor search and grid index is suitable for line segment organization. So, we design a hierarchical grid index with multiple resolutions to process trajectory segments of different lengths, and propose a non-trivial search algorithm to find the $K$-nearest neighbours for finishing the trajectory modification tasks efficiently. For simplicity, we mainly focus on intra-trajectory modification in this section because: 1) The global utility loss at trajectory level is aggregated from the local one at trajectory-segment level, and hence improving the efficiency of intra-trajectory modification would naturally speed up inter-trajectory modification; 
2) The overall search procedure is essentially similar, thus we simply use single trajectory modification to illustrate the main idea of our solutions. We elaborate on the index structure and search algorithm in the following.

\subsubsection{\textbf{Index Structure}}
Let $\mathcal{G}_r$ denote a set of grid cells in a uniform grid of granularity $r$ (e.g., $512\times 512$ cells). The hierarchical grids $\mathcal{HG}=\{\mathcal{G}_{r_1},\dots, \mathcal{G}_{r_H}\}$ consist of multiple-level grids with different granularity, where $r_1<\dots<r_H$. Usually, $\mathcal{G}_{r_1}$ is the coarsest-grained grid representing the entire spatial area with the granularity of $1\times 1$ and the finest granularity $r_H$ depends on the density of trajectory points. 

Regarding each single grid cell $g_r^i$ in $\mathcal{G}_r$, three pieces of information will be recorded: 1) the geographic coverage; 2) all trajectory segments that entirely locate in $g_r^i$ but cannot further fit in any finer grid cell $g_{r^\prime}^j$ in our hierarchical grids $\mathcal{HG}$, where $r^\prime>r$; and 3) the parent/children pointers encoding the hierarchical relationship (as exemplified in \Cref{fig:eg-grid}). Specifically, we define the parent and children of an arbitrary grid cell $g_r^i$ in $\mathcal{G}_r$ as follows:
\begin{itemize}
    \item Its parent should be a coarser grid cell in $\mathcal{G}_{r^\prime}$ (i.e., $r^\prime<r$) and be the smallest one that can completely enclose $g_r^i$.
    \item Its children should be a set of finer grid cells in $\mathcal{G}_{r''}$ where $r''>r$, each of which entirely locates in $g_r^i$ but cannot be covered by any other finer-grained grid cell than $g_r^i$. 
\end{itemize}

Meanwhile, we assign every possible trajectory segment to a specific grid cell based on the following criteria:
\begin{definition}[Best-fit Grid Cell]
    Given a trajectory segment $s=\langle p_x, p_y\rangle$, a grid cell $g_r^i$ in the grid $\mathcal{G}_r$ is called the ``best-fit'' grid cell of $s$, if and only if: 
    \begin{itemize}
        \item the two endpoints $p_x$ and $p_y$ locate in the same grid cell $g_r^i$ in $\mathcal{G}_r$, namely, $p_x\in g_r^i \wedge p_y\in g_r^i$ where $g_r^i\in \mathcal{G}_r$;
        \item they locate in two different grid cells in a finer granularity $\mathcal{G}_{r^\prime}$ with $r^\prime>r$, namely, $p_x\in g_{r^\prime}^j \wedge p_y\in g_{r^\prime}^k$ where $j\neq k$. 
    \end{itemize}
\end{definition}
As illustrated in \Cref{fig:eg-grid}, the best-fit grid cells for trajectory segments $\langle p_1, p_2\rangle$ and $\langle p_5, p_6\rangle$ are $g_1^1$ and $g_2^4$ respectively.
\vspace{2mm}

\subsubsection{\textbf{Bottom-Up-Down Search}}\label{sec:searchalgo}
We first define a novel distance metric that will be utilized in our pruning strategy for $K$-nearest trajectory segment search, and provide a proof of the correctness of the pruning method. We then introduce our search algorithm designed for efficient trajectory modification.

\begin{definition}[Point-Grid Cell Distance]
    To compute the minimum distance from a point $q$ to a grid cell $g$ that is a rectangle with four connected edges, we first check whether $q$ locates inside $g$. If so, then $MINdist(q,g)=0$; otherwise, it depends on the edge $\bar{s}$ of $g$ which is the closest to $q$:
    \begin{equation}
        \label{eq:gridd}
        MINdist(q,g) = \left\{ 
        \begin{array}{ll}
        \min_{\bar{s}\in g} dist(q,\bar{s}) & \textrm{if $q\notin g$}\\
        0 & \textrm{if $q\in g$}
        \end{array} \right.
    \end{equation}
\end{definition}
The minimum distance between $q$ and a candidate grid cell can help us to prune unpromising children cells and containing trajectory segments based on the following theorem:

\begin{theorem}
    \label{theorem:search}
    Given a grid cell $g$ in $\mathcal{G}_r$, a point $q$ and its current $K$-th closest segment $s_K$, if the minimum distance from $q$ to $g$ is greater than $\theta_K=dist(q, s_K)$, namely, $MINdist(q,g)>\theta_K$, then all trajectory segments within $g$ and its children (i.e., all finer-grained grid cells inside $g$ in $\mathcal{G}_{r'}$ with $r'>r$) will not be the promising candidates for $q$'s $K$-nearest neighbours.
\end{theorem}
\begin{proof}
   Any segment $s$ inside $g$ satisfies the inequality property that $dist(q,s)\geq MINdist(q,g)$ based on the Point-Segment Distance in \Cref{eq:segd} and Point-Grid Cell Distance in \Cref{eq:gridd}. Therefore, $\nexists s\in g, \textrm{s.t.}~dist(q,s)\leq \theta_K$. Similarly, no finer-grained cell in $\mathcal{G}_{r'}$ with $r'>r$ can contribute a segment whose distance to $q$ is smaller than $\theta_K$. Hence, grid cell $g$ and all its children can be pruned safely for $K$-NN search.
\end{proof}

Typically, the top-down or best-first algorithm is utilized when using a hierarchical index structure to speed up the kNN search. However, in our $K$-nearest trajectory (segment) search for a point $q$, we notably observe that some unpromising segments which locate in promising grid cells will be checked earlier than the final top-$K$ candidates, if performing the search on the hierarchical grids in a straightforward top-down manner. We illustrate an example as follows:

\begin{figure}[hbt]
    \flushleft
    \begin{tabular}{rl}
        \hspace*{-2mm}\includegraphics[height=0.2\textwidth]{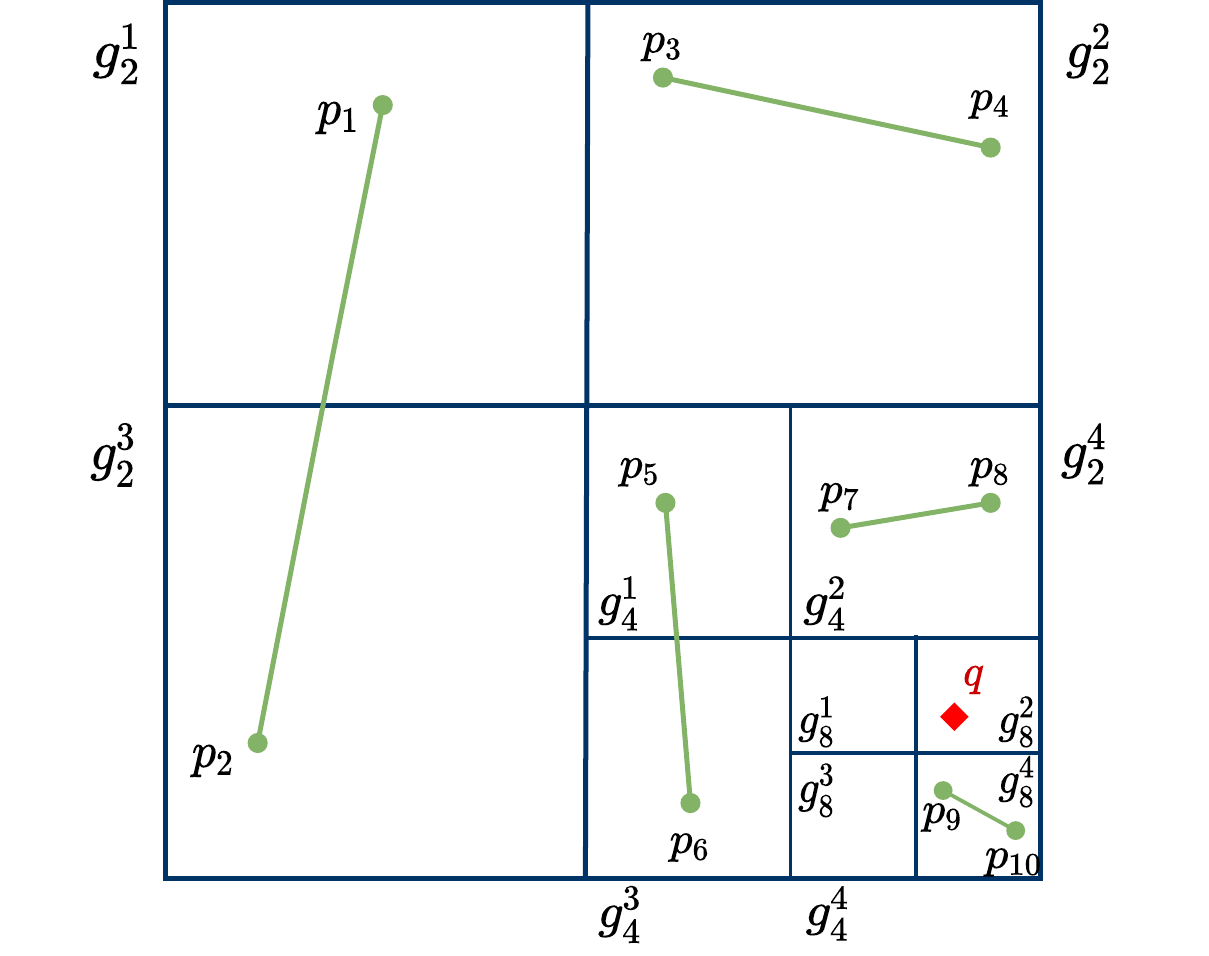}
        &
        \hspace*{-4mm}\includegraphics[height=0.18\textwidth]{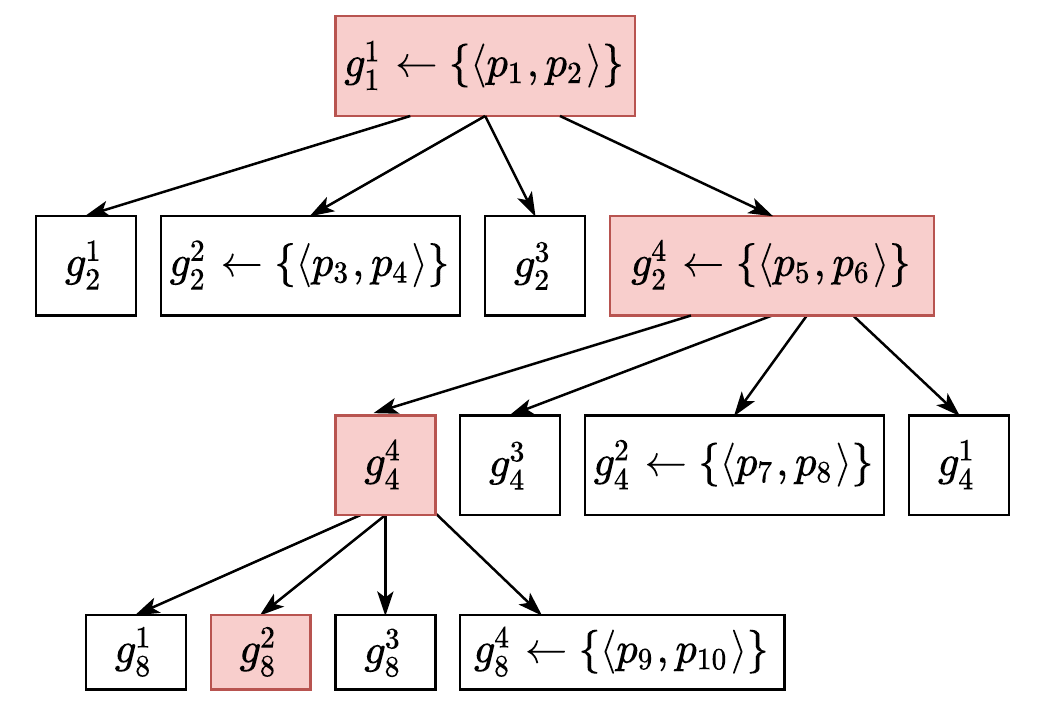}
    \end{tabular}
    \caption{An example of segment-based hierarchical grid index.}
    \label{fig:eg-grid}
\end{figure}

\begin{example}
    The entire area, namely the only grid cell in the coarsest grid $\mathcal{G}_1$, is denoted as $g_1^1$ and contains the segment $\langle p_1, p_2\rangle$. Other segments including $\langle p_3, p_4\rangle$, $\langle p_5, p_6\rangle$, $\langle p_7, p_8\rangle$ and $\langle p_9, p_{10}\rangle$ locate in finer grid cells $g_2^2$, $g_2^4$, $g_4^2$ and $g_8^4$, respectively. Assume we are looking for the top-1 nearest segment for the query point $q$ in $g_8^2$ (in red). Apparently, the segment $\langle p_9,p_{10}\rangle$ is the target one with the minimum distance. If such distance can be identified earlier and utilized as the threshold, it would help to safely prune $g_2^2$ and $g_4^2$ without computing the exact distance between $q$ and the inside segments. However, traditional top-down search checks $\langle p_1,p_2\rangle$ and $\langle p_5,p_6\rangle$ first (as they locate in promising grids with zero MINdist to $q$), which only produces a very loose distance threshold, resulting in unnecessary calculation of $\langle p_3,p_4\rangle$ and $\langle p_7,p_8\rangle$ before the target $\langle p_9,p_{10}\rangle$ is found.  
\end{example}

To avoid such unnecessary calculation, we design a novel \textit{Bottom-Up-Down} search algorithm to obtain the $K$-nearest trajectory segments more efficiently. Without losing generality, we here explain the nearest segment search (i.e., searching the top-1 result) for point insertion as an example. 

\begin{algorithm}[htb]
    \small
    \caption{$K$-Nearest Segment Search}
    \label{algo:local_insert}
    \KwIn{$\mathcal{HG}$: the hierarchical grids; $\tau$: target trajectory; $q$: target point to be inserted; $\Delta f_q$: number of insertions}
	\KwOut{$\tau^*$: the modified trajectory}
    $g_{r_H}\leftarrow$ LocatePoint$(q, \mathcal{HG}(\tau), r_H)$\tcp*[r]{start from finest grid}
    $Q_s\leftarrow \emptyset$\tcp*[r]{priority queue for candidate segments}
    $S_g\leftarrow \emptyset$\tcp*[r]{stack for grid cells used in bottom-up}
    $S_g.push(\langle g_{r_H},0\rangle)$\tcp*[r]{$MINdist(q,g_{r_H})=0$}
    $Q_g\leftarrow \emptyset$\tcp*[r]{priority queue for grid cells in top-down}
    $V\leftarrow \emptyset$\tcp*[r]{record all visited grid cells}
    $\theta_K \leftarrow +\infty$;~~
    $rootAccess\leftarrow false$;\\
    \While{$|S_g|>0 \vee |Q_g|>0$}{
        \If{$rootAccess=false$}{
            $\langle g_{candi}, dist\rangle \leftarrow S_g.pop()$\tcp*[r]{pop from stack}
            \If{$|Q_s|\geq \Delta f_q \wedge dist > \theta_K$}{
                continue;
            }
        }
        \Else{
            $\langle g_{candi}, dist\rangle \leftarrow Q_g.pop()$\tcp*[r]{pop from queue}
            \If{$|Q_s|\geq \Delta f_q \wedge dist > \theta_K$}{
                break\tcp*[r]{earlier termination if possible}
            }
        }
        $V\leftarrow V\cup candi_g$\tcp*[r]{mark it as visited}
        
        \tcc{check all segments within the grid cell $g_{candi}$}
        \For{$\forall s\in g_{candi}$}{
            $dist \leftarrow$ ComputeDistance$(q,s)$; \\
            \If{$|Q_s|<\Delta f_q$ or $dist < \theta_K$}{
                $Q_s.push(\langle s, dist\rangle)$;
            }
        }
        \If{$|Q_s|\geq\Delta f_q$}{
            $\theta_K\leftarrow Q_s.top().dist$\tcp*[r]{update the threshold}
        }
        \tcc{push unvisited candidates into the stack or queue}
        \If{$rootAccess=false \wedge g_{candi}.parent \not\in V$} {
            \If{$g_{candi}.parent\in\mathcal{G}_{r_1}$}{
                $rootAccess=true$;\\
                $Q_g.push(\langle g_{candi}.parent, 0\rangle)$\tcp*[r]{$Q_g$ replaces $S_g$}
            }
            \Else{
                $S_g.push(\langle g_{candi}.parent,0\rangle)$;\\
            }
        }
        
        \tcc{check all children grid cells of $candi_g$}
        \For{$g_c \in g_{candi}.children \wedge g_c \not\in V$}{
            $mindist\leftarrow$ ComputeMINdist$(q,g_c)$;\\
            \If{$rootAccess=false$}{
                $S_g.push(\langle g_c,mindist\rangle)$;
            }
            \Else{
                $Q_g.push(\langle g_c,mindist\rangle)$;
            }
        }
    }
    $\tau^*\leftarrow$ ModifyAndUpdate$(\tau,\mathcal{HG(\tau)},Q_s,q)$;\\
    \Return{$\tau^*$};
\end{algorithm}

Intuitively, the closest segment is more likely to locate in a finer grid cell nearby the query point, even though they do not stay in exactly the same grid cell. Starting from the finest-grained grid (i.e., $\mathcal{G}_{r_H}$) can help to quickly obtain a relatively smaller distance threshold $\theta_K$ in an earlier stage, which is admittedly capable of providing more powerful pruning without damaging the search correctness. Therefore, we first conduct a ``bottom-up'' search starting from the finest-grained grid cell where the query point $q$ locates (line 1 in \Cref{algo:local_insert}). When processing a grid cell $g_{candi}$, all segments within $g_{candi}$ are checked and maintained by a priority queue (lines 18-21). If a promising candidate is identified, we tighten the distance threshold $\theta_K$ to enhance pruning (line 23). Meanwhile, the parent of $g_{candi}$ and its unvisited children will be considered as well (lines 24-35). Note that we first push the parent and then the children into the stack in order to check the more promising finer-grained grid cells earlier even though the $MINdist$ to the point $q$ is non-zero. We repeat this bottom-up checking process until the root node is reached (line 26), implying all grid cells with zero $MINdist$ to $q$ have been visited already and a tight distance bound has been achieved. Thus, we move to a ``top-down'' search manner via the grid priority queue $Q_g$ whose priority is the non-zero $MINdist$ to $q$. Again, it helps to terminate the search as long as the currently top-1 candidate grid cell is worse than the lower bound $\theta_K$ (line 14-16), which means it is impossible to find any segment or grid cell with a smaller distance to $q$ (\Cref{theorem:search}). Finally, we insert point $q$ into these returned candidate segments as the anonymized trace $\tau^*$ and update the hierarchical grids $\mathcal{HG}(\tau)$ accordingly.

\section{Experiments}\label{sec:exp}
We have conducted extensive experiments on a real-world trajectory dataset to evaluate the effectiveness and efficiency of our proposed privacy model. The detailed results are reported in this section. All the algorithms are implemented in C++, and all the experiments are run on a server with two Intel(R) Xeon(R) CPU E5-2630, 10 cores/20 threads at 2.2GHz each, 378GB memory, and Ubuntu 16.04 operating system.

\subsection{Experimental Setting}
\noindent\textbf{Dataset:}
We evaluate our proposed algorithms on a publicly-available one-week taxi dataset, \textit{T-Drive} \cite{yuan2010tdrive}, generated by 10,357 Beijing taxis with more than 15 million GPS points in total. The Euclidean distance between two continuous points is around 600 meters and the average sampling rate is 3.1 minutes per point. Note that each taxi is associated with a single trajectory covering its entire moving traces and every trajectory consists of 1,813 points on average.

\noindent\textbf{Compared Methods and Parameter Setting:}
We compare our algorithms with several leading privacy models, and the parameters are set based on the original work:
\begin{itemize}
    \item \textit{K-anonymity-based models:} 1) W4M\cite{abul2010anonymization} ensures each trajectory is indistinguishable with other k-1 trajectories in a cluster; 2) GLOVE\cite{gramagliaF15hiding} achieves k-anonymity for region-based trajectories via spatiotemporal generalization; 3) KLT\cite{tu2019protecting} extends GLOVE by ensuring not only k-anonymity but also l-diversity and t-closeness ($k=5$ for all models and $l=3,t=0.1$ for KLT);
    \item \textit{Signature closure} (SC)\cite{jin2020trajectory} is an effective trajectory protection method by discarding all top-$m$ signature points ($m=10$); \textit{Radius-based signature closure} (RSC-$\alpha$) is an enhanced variant of SC dropping extra points within a radius $\alpha$ centered at signature points ($\alpha\in[0.1,0.5,1,3,5]$);
    \item \textit{Generation-based DP models:} 1) DPT\cite{he2015dpt} is a pioneering model generating differentially private synthetic trajectories; 2) AdaTrace\cite{gursoy2018utility,DBLP:journals/tmc/Gursoy0TY19} outperforms DPT by combing differential privacy with attack resilience and utility-aware generator ($\epsilon=1.0$);
    \item \textit{Frequency-based randomized DP models:} 1) \textit{PureG} denotes our global mechanism with $\epsilon_G$ privacy; 2) \textit{PureL} represents our local mechanism with $\epsilon_L$ privacy; 3) \textit{GL} combines our global and local mechanisms (with exchangeable composition ordering) providing $\epsilon$-differential privacy. ($\epsilon_G=\epsilon_L=0.5$ and $\epsilon=\epsilon_G+\epsilon_L=1.0$).
\end{itemize}

\noindent\textbf{Evaluation Metrics:}
We compare these models in terms of privacy protection, utility preservation and data recovery.
\begin{itemize}
    \item \textit{Privacy protection}: linking accuracy (LA) by a state-of-the-art user re-identification model \cite{jin2019moving} - smaller LA represents better privacy protection. In specific, we employ all types of signatures provided in \cite{jin2019moving} which capture different features of trajectories: $LA_s$ (via spatial signature);  $LA_t$ (via temporal signature);  $LA_{st}$ (via spatiotemporal signature); $LA_{sq}$ (via sequential signature); mutual information (MI) \cite{DBLP:conf/ccs/YangLQQMM12,li2017achieving} generally measures the dependency between the original and anonymized data 
    - smaller MI means better protection.
    \item \textit{Utility preservation}: point-based information loss (INF) \cite{han2015sst} at the statistical level; the divergence of diameter distribution (DE) and trip distribution (TE) \cite{gursoy2018differentially} from the spatial aspect; and the F-measure of frequent pattern mining (FFP) \cite{gurung2014traffic} - smaller INF, DE, TE and larger FFP indicate better utility preservation.
    \item \textit{Data Recovery}: the success rate of trajectory recovery via an HMM-based map-matching technique measured from various aspects \cite{newson2009hmm,chao2020survey}. 
\end{itemize}

We also evaluate model efficiency by comparing the naive \textit{linear scan} approach and the single-level uniform grid index ($UG$) to our hierarchical grid index with various kNN search strategies: \textit{top-down} ($HG_t$), \textit{bottom-up} ($HG_b$) and \textit{bottom-up-down} ($HG_+$ explained in \Cref{sec:searchalgo}). The granularity of uniform grid and the finest level in our hierarchical grid index is empirically set to $512\times 512$ cells.

\subsection{Effectiveness Evaluation}
\Cref{tab:summary} reports the empirical results about the effectiveness comparison among the anonymizaion models.
\vspace{0.5mm}

\begin{table*}[htb]
\scriptsize
\renewcommand{\arraystretch}{1.1}
\centering
\caption{A summary of effectiveness evaluation results ($|D|=1000$ and $\epsilon=1.0$)}
\label{tab:summary}
\begin{tabular}{c|c|cccccc|ccc|cc|ccc}
\hline
                          & Metric    & SC    & RSC-0.1 & RSC-0.5 & RSC-1.0 & RSC-3.0 & RSC-5.0 & W4M   & GLOVE & KLT   & DPT    & AdaTrace & PureG & PureL & GL    \\\hline
\multirow{5}{*}{Privacy}  & LA$_s~$    & 0.188 & 0.146   & 0.098   & 0.081   & 0.049   & 0.033   & 0.847 & 0.370 & 0.269 & 0.006 & 0.000    & 0.922 & 0.119 & 0.016 \\
                          & LA$_t~$    & 0.206 & 0.253   & 0.225   & 0.122   & 0.068   & 0.024   & 0.164 & 0.005 & 0.002 & -     & -        & 0.211 & 0.028 & 0.022 \\
                          & LA$_{st}$ & 0.888 & 0.889   & 0.869   & 0.846   & 0.750   & 0.584   & 0.506 & 0.116 & 0.099 & -      & -        & 0.928 & 0.362 & 0.337 \\
                          & LA$_{sq}$ & 0.206 & 0.129   & 0.098   & 0.075   & 0.043   & 0.032   & 0.860 & 0.350 & 0.324 & 0.009  & 0.000    & 0.975 & 0.026 & 0.015 \\
                          & MI        & 0.163 & 0.162   & 0.158   & 0.156   & 0.150   & 0.148   & 0.244 & 0.385 & 0.312 & 0.097  & 0.057    & 0.184 & 0.098 & 0.095 \\\hline
\multirow{4}{*}{Utility}  & INF       & 0.656 & 0.663   & 0.682   & 0.704   & 0.784   & 0.853   & 0.285 & 0.912 & 0.929 & 0.986  & 0.603    & 0.495 & 0.635 & 0.642 \\
                          & DE        & 0.047 & 0.050   & 0.064   & 0.082   & 0.138   & 0.243   & 0.057 & 0.587 & 0.617 & 0.297  & 0.291    & 0.004 & 0.015 & 0.014 \\
                          & TE        & 0.385 & 0.380   & 0.373   & 0.370   & 0.333   & 0.308   & 0.330 & 0.657 & 0.686 & 0.375  & 0.064    & 0.225 & 0.365 & 0.331 \\
                          & FFP       & 0.990 & 0.988   & 0.983   & 0.977   & 0.943   & 0.929   & 0.994 & 0.416 & 0.330 & 0.373  & 0.908    & 0.980 & 0.953 & 0.956 \\\hline
\multirow{5}{*}{Recovery} & Precision & 0.610 & 0.610   & 0.606   & 0.598   & 0.561   & 0.535   & 0.276 & 0.498 & 0.435 & -      & -        & 0.378 & 0.307 & 0.309 \\
                          & Recall    & 0.611 & 0.605   & 0.578   & 0.540   & 0.384   & 0.251   & 0.157 & 0.307 & 0.276 & -      & -        & 0.481 & 0.331 & 0.339 \\
                          & F-score   & 0.611 & 0.608   & 0.591   & 0.568   & 0.456   & 0.342   & 0.179 & 0.380 & 0.338 & -      & -        & 0.423 & 0.318 & 0.324 \\
                          & RMF       & 0.779 & 0.782   & 0.798   & 0.823   & 0.916   & 0.967   & 0.693 & 0.516 & 0.592 & -      & -        & 1.310 & 1.342 & 1.420 \\
                          & Accuracy  & 0.162 & 0.162   & 0.162   & 0.162   & 0.160   & 0.152   & 0.035 & 0.250 & 0.218 & -      & -        & 0.395 & 0.189 & 0.008 \\\hline
\end{tabular}
\begin{tablenotes}
\item $^1$~We ignore the results of LA$_t$ and LA$_{st}$ for \textit{DPT} and \textit{AdaTrace} since both models generate synthetic trajectories only in the spatial dimension without any timestamps.
\item $^2$~We ignore the data recovery experiments for \textit{DPT} and \textit{AdaTrace} since the generated synthetic trajectories are no longer aligned to the road network.
\end{tablenotes}
\end{table*}
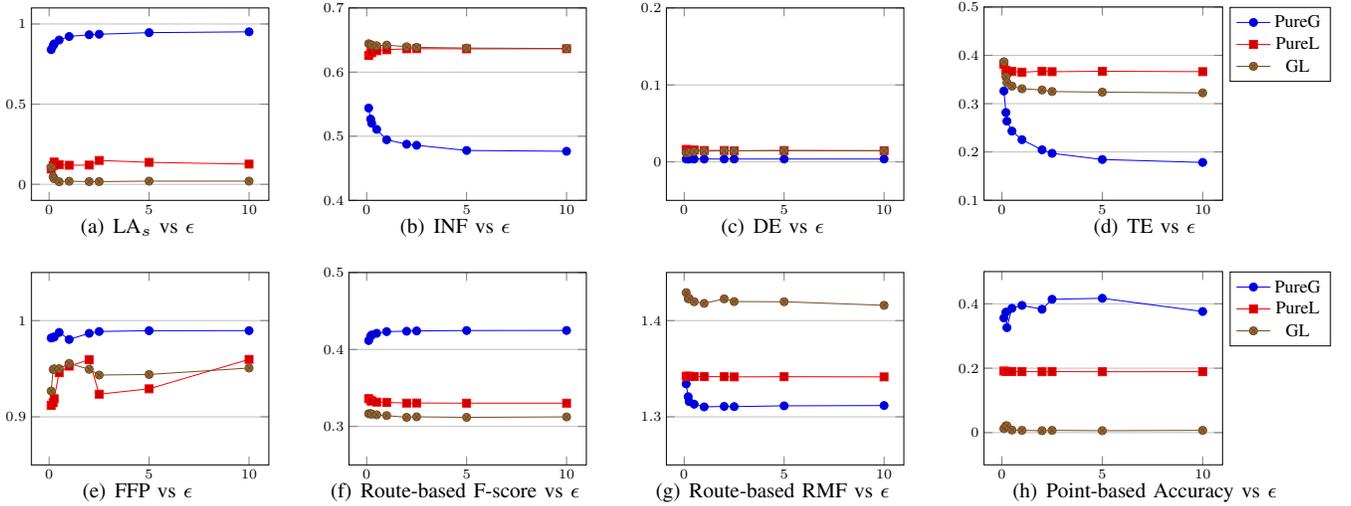
\begin{figure*}[htb]
    \centering
    \scriptsize
    \begin{tabular}{rrrr}
    \subfigure[LA$_s$ vs $\epsilon$]{
    \resizebox{0.2\textwidth}{!}{%
    \begin{tikzpicture}
        \begin{axis}[
        height=5cm,
        ymajorgrids,
        ymin=-0.1,
        ymax=1.1,
        legend pos=outer north east
        ]
            \addplot table[x=EPS,y=pureG] {tables/linking.csv};
            \addplot table[x=EPS,y=pureL] {tables/linking.csv};
            \addplot table[x=EPS,y=GL] {tables/linking.csv};
        \end{axis}
    \end{tikzpicture}
    }
    } &
    \subfigure[INF vs $\epsilon$] {
    \resizebox{0.2\textwidth}{!}{%
    \begin{tikzpicture}
        \begin{axis}[
        height=5cm,
        ymajorgrids,
        ymin=0.4,
        ymax=0.7
        ]
            \addplot table[x=EPS,y=pureG] {tables/utility_infoloss.csv};
            \addplot table[x=EPS,y=pureL] {tables/utility_infoloss.csv};
            \addplot table[x=EPS,y=GL] {tables/utility_infoloss.csv};
        \end{axis}
    \end{tikzpicture}
    }
    } &
    \subfigure[DE vs $\epsilon$]{
    \resizebox{0.2\textwidth}{!}{%
    \begin{tikzpicture}
        \begin{axis}[
        height=5cm,
        ymajorgrids,
        ymin=-0.05,
        ymax=0.2,
        ytick={0,0.1,0.2},
        ]
            \addplot table[x=EPS,y=pureG] {tables/utility_de.csv};
            \addplot table[x=EPS,y=pureL] {tables/utility_de.csv};
            \addplot table[x=EPS,y=GL] {tables/utility_de.csv};
        \end{axis}
    \end{tikzpicture}
    }
    } &
    \subfigure[TE vs $\epsilon$]{
    \resizebox{0.28\textwidth}{!}{%
    \begin{tikzpicture}
        \begin{axis}[
        height=5cm,
        ymajorgrids,
        ymin=0.1,
        ymax=0.5,
        legend pos=outer north east
        ]
            \addplot table[x=EPS,y=pureG] {tables/utility_te.csv};\addlegendentry{\small PureG}
            \addplot table[x=EPS,y=pureL] {tables/utility_te.csv};\addlegendentry{\small PureL}
            \addplot table[x=EPS,y=GL] {tables/utility_te.csv};\addlegendentry{\small GL}
        \end{axis}
    \end{tikzpicture}
    }
    } \\
    \subfigure[FFP vs $\epsilon$]{
    \resizebox{0.2\textwidth}{!}{%
    \begin{tikzpicture}
        \begin{axis}[
        height=5cm,
        ymajorgrids,
        ymin=0.85,
        ymax=1.05,
        ytick={0.9,1.0},
        ]
            \addplot table[x=EPS,y=pureG] {tables/utility_ffp.csv};
            \addplot table[x=EPS,y=pureL] {tables/utility_ffp.csv};
            \addplot table[x=EPS,y=GL] {tables/utility_ffp.csv};
        \end{axis}
    \end{tikzpicture}
    }
    } &
    \subfigure[Route-based F-score vs $\epsilon$]{
    \resizebox{0.2\textwidth}{!}{%
    \begin{tikzpicture}
        \begin{axis}[
        height=5cm,
        ymajorgrids,
        ymin=0.25,
        ymax=0.5,
        ]
            \addplot table[x=EPS,y=pureG] {tables/recovery_routeF.csv};
            \addplot table[x=EPS,y=pureL] {tables/recovery_routeF.csv};
            \addplot table[x=EPS,y=GL] {tables/recovery_routeF.csv};
        \end{axis}
    \end{tikzpicture}
    }
    } &
    \subfigure[Route-based RMF vs $\epsilon$] {
    \resizebox{0.2\textwidth}{!}{%
    \begin{tikzpicture}
        \begin{axis}[
        height=5cm,
        ytick={1.3,1.4},
        ymajorgrids,
        ymin=1.25,
        ymax=1.45,
        ]
            \addplot table[x=EPS,y=pureG] {tables/recovery_routeRMF.csv};
            \addplot table[x=EPS,y=pureL] {tables/recovery_routeRMF.csv};
            \addplot table[x=EPS,y=GL] {tables/recovery_routeRMF.csv};
        \end{axis}
    \end{tikzpicture}
    }
    } &
    \subfigure[Point-based Accuracy vs $\epsilon$] {
    \resizebox{0.28\textwidth}{!}{%
    \begin{tikzpicture}
        \begin{axis}[
        height=5cm,
        ymajorgrids,
        ymin=-0.1,
        ymax=0.5,
        legend pos=outer north east,
        ]
            \addplot table[x=EPS,y=pureG] {tables/recovery_point.csv};\addlegendentry{\small PureG}
            \addplot table[x=EPS,y=pureL] {tables/recovery_point.csv};\addlegendentry{\small PureL}
            \addplot table[x=EPS,y=GL] {tables/recovery_point.csv};\addlegendentry{\small GL}
        \end{axis}
    \end{tikzpicture}
    }
    } \\
    \end{tabular}
    \caption{The impact of $\epsilon$ ($|D|=1000$)}
    \label{fig:epsilon}
    \vspace{-1mm}
\end{figure*}

\subsubsection{\textbf{Privacy Protection}}
Overall, k-anonymity-based models are quite ineffective for defending against the powerful linking attack, although the mutual information between the original and the anonymized trajectories is acceptable to some extent. 
It is not surprising that \textit{GLOVE} and \textit{KLT} can perform well when using temporal-related signatures ($LA_t<0.1\%$ and $LA_{st}\approx 10\%$) mainly due to their brute generalization - each sample is generalized into a region and a time range rather than an accurate spatiotemporal point. On the contrary, \textit{DPT} and \textit{AdaTrace}, as remarkable generative differential privacy models, are the most effective in privacy protection but with a huge sacrifice in data utility. \textit{SC} also achieves relatively low linkage accuracy via all types of signatures except the spatiotemporal one. Similar trend happens in \textit{RSC} which extends the point removal to a region centred at the signature points. This verifies the importance of signature points to prevent re-identification attack.
Our DP model introduces frequency-based randomization on the signatures. It is reasonable to doubt its capability against powerful linkage attacks, since the occurrence of some essential points might be randomly increased after the noise injection. We admit that the \textit{PureG} model performs unsatisfactorily overall, despite globally perturbing trajectory frequency distribution is capable of damaging some mutual information. Fortunately, simply implementing the local point frequency perturbation and intra-trajectory modification (i.e., \textit{PureL}) can dramatically reduce the linkage accuracy already ($LA_{st}\approx 36\%$ and $LA_s<12\%$). Moreover, the effect of protection can be further magnified when both local and global mechanisms are integrated based on the ``composition theorem'' in \Cref{dp:composition}, especially when dealing with the spatial-based linking attack ($LA_s=1.6\%$). This undoubtedly verifies the effectiveness of our proposed DP model in privacy protection.
It is worth noting that although our privacy model only alters the local/global frequency distribution of spatial signature points while ignoring the temporal information, it naturally changes the other types of signatures as well and thus gain further protection. As demonstrated in \Cref{tab:summary}, the linking accuracy using temporal signature ($LA_t$), spatiotemporal signature ($LA_{st}$) and sequential signature ($LA_{sq}$) also achieves significant decrease after applying our \textit{GL} model on the original data.

\vspace{0.5mm}
\subsubsection{\textbf{Utility Preservation}}
Ideally, privacy protection should not be achieved at the cost of huge utility loss. \textit{DPT} performs the worst in utility preservation among all the compared DP-based methods. Although $DE$ and $TE$ are still relatively low ($\approx 30\%$) as \textit{DPT} utilizes prefix trees to capture the objects' spatial transition behaviour, it ends up with a low $FFP$ and an extremely high $INF$. In particular, $INF$ measures how many original points are lost after the anonymization. \textit{DPT} destroys almost $99\%$ of the points due to the transmit-based synthetic generation framework, which apparently impacts the performance of frequent pattern mining. \textit{AdaTrace} shows better ability to reserve more data utility thanks to its specially designed utility-aware synthesizer superior to \textit{DPT}. In comparison, k-anonymity based approaches perform differently: \textit{W4M} is absolutely the best k-anonymity model to make the anonymized data sufficiently useful and, as mentioned, \textit{GLOVE} and \textit{KLT} outperform \textit{W4M} in privacy protection which is at the price of extensive utility loss.
On the contrary, the signature-based methods (\textit{SC}, \textit{RSC}, and ours) distort only a limited number of signature points while keeping most of the other points unchanged, which naturally achieves satisfactory performance in utility preservation. The \textit{RSC} variants with different radius $\alpha$ are surpassed by \textit{SC}, with more points being deleted from the original trajectories. From \Cref{tab:summary} we can observe that all the three versions of our proposed DP model reserve sufficient data utility after frequency perturbation and trajectory modification ($INF=60\%$, $TE=30\%$ and $FFP=96\%$). In particular, our model is extraordinarily powerful in retaining the diameter information in the anonymized trajectories, as demonstrated by a negligible divergence of diameter distribution ($DE<1.5\%$). Overall, randomizing the frequency distribution of trajectory points instead of brutally discarding them can provide not only the strong capability of resisting various privacy attacks but also the controlled data utility under the differential privacy guarantee, outperforming the generation-based DP models.

\subsubsection{\textbf{Data Recovery}}
Trajectory privacy has been well-studied recently, while few works have paid attention to the risk of recovery attack, i.e., the capability of reconstructing the original data from the anonymized one using some map-matching or path inference technologies. In our experimental study, we use the well-known HMM-based map-matching algorithm \cite{newson2009hmm} to simulate the recovery attack on the anonymized trajectory dataset, and apply various metrics to evaluate the success rate of data recovery. Specifically, the \textit{Precision/Recall/F1-score} and the length-based \textit{Route Mismatch Fraction} ($RMF$) evaluate from the perspective of route-based matching \cite{wang2014RMF}, whilst the \textit{Accuracy} shows the performance of point-based matching \cite{chao2020survey}. Note that our frequency-based randomization mechanisms might inject more points into the original dataset, making the output trajectories longer. Thus, it is likely that $RMF$ exceeds 1; the higher the $RMF$, the more erroneous the recovery and the better protection the model offers. We ignore the recovery results for \textit{DPT} and \textit{AdaTrace} as they are generative models that completely synthesize trajectories.

Although signatures are vital to spatiotemporal entity linking and trajectory privacy protection \cite{jin2019moving,jin2020trajectory}, simply removing these personally-identifying signature points (\textit{SC}) and their nearby neighbours (\textit{RSC}) is insufficient for trajectory privacy, since many raw points ($>60\%$ in \Cref{tab:summary}) can be recovered by the HMM-based map-matching. Interestingly, recovering the trajectories anonymized by \textit{W4M} is much more difficult, since \textit{W4M} enforces that each trajectory should be spatially close with its pivot trajectory, making itself deviate from the real paths. The region-based generalized trajectories from \textit{GLOVE} and \textit{KLT} are highly possible to hit one of its k-components so as to be successfully recovered via powerful map-matching. By contrast, our frequency-based differential privacy models (i.e., \textit{PureG}, \textit{PureL}, and \textit{GL} with exchangeable ordering of local and global mechanisms) are capable to well resisting recovery attacks, as evidenced by the poor map-matching results ($\approx 33\%$ route-based Precision/Recall/F-score and $\approx 10\%$ point-based Accuracy). It again proves the effectiveness of our main idea that perturbing the frequency distributions of trajectory points rather than the location coordinates can help to protect the most sensitive geo-information. In other words, introducing probabilistic noise to the distributions leads to an unpredictable raise or drop of the point frequency, which is hard to recover from the anonymized trajectories.

\subsubsection{\textbf{The Impact of $\epsilon$ and the Comparison Between Global and Local Mechanisms}}
Two independent DP mechanisms, namely, global TF perturbation over the entire dataset and local PF perturbation for each trajectory, have been proposed. According to \Cref{dp:composition}, we combine them together to gain more privacy guarantee $\epsilon=\epsilon_G+\epsilon_L$ by evenly allocating the total privacy budget, i.e., $\epsilon_G=\epsilon_L=\frac{1}{2}\epsilon$. In this section, we investigate the impact of the privacy budget on the overall performance. Specifically, we set $\epsilon_G$ and $\epsilon_L$ within $[0.1,10.0]$ and report the effectiveness of the three variants (namely, \textit{PureG}, \textit{PureL}, and \textit{GL}) in \Cref{fig:epsilon}.

It can be observed that simply utilizing the global mechanism (\textit{PureG}) performs the worst in balancing privacy protection and utility preservation, and the gap is gradually enlarged (especially in $LA_s$) when the privacy budget $\epsilon$ grows. Considering the anonymization power, \textit{PureG} indeed obscures the unique information in the whole dataset, i.e., some distinctive locations visited by fewer users. However, the sensitive points that are highly representative of a specific individual have not been distorted properly. On the other hand, the perturbation of local frequency distribution hides most of the personally-identifying information from the trajectories with a novel Laplace noise injection mechanism, achieving satisfactory protection. Whilst the inferiority of \textit{PureG} in preventing attacks is compensated by the improved data utility, as verified by lower $INF$ and $TE$ as well as higher $FFP$. Furthermore, the global alteration is much easier to be recovered by map-matching than the local alteration as less changes are made to each trajectory during the inter-trajectory modification.
Overall, utility preservation enhances while privacy protection degrades with the increase of privacy budget, mainly because less noises are injected into the trajectory data from Laplace distribution when $\epsilon$ raises. Our frequency-based randomization mechanisms (in particular the models that integrate local PF alteration) are not quite sensitive to the change of $\epsilon$, as the trend in \Cref{fig:epsilon} is relatively stable.

\subsection{Efficiency Evaluation}
\begin{figure}
    \begin{tabular}{ll}
    \subfigure{ 
    \resizebox{0.23\textwidth}{!}{%
    \begin{tikzpicture}
    \begin{axis}
    [
    ymode=log,
    ymajorgrids=true,
    ylabel={\footnotesize Time costs ($\log$)},
    ybar,
    /pgf/bar width=3.3pt,
    legend style={at={(0.5,-0.3)}, 
    anchor=south,legend columns=-1},   
    symbolic x coords={{1000},{2000},{4000},{6000},{8000},{10000}},
    xtick=data,
    ]
        \addplot  table[x=D,y=Linear] {tables/timecost3.csv}; \addlegendentry{\small Linear}
        \addplot  table[x=D,y=UG] {tables/timecost3.csv}; \addlegendentry{\small $UG$}
        \addplot table[x=D,y=HG_t] {tables/timecost3.csv}; \addlegendentry{\small $HG_t$}
        \addplot  table[x=D,y=HG_b] {tables/timecost3.csv};\addlegendentry{\small $HG_b$}
        \addplot  table[x=D,y=HG] {tables/timecost3.csv}; \addlegendentry{\small $HG_+$}
        \legend{Linear, $UG$, $HG_t$, $HG_b$, $HG_+$}
    \end{axis}
    \end{tikzpicture}
    }
    } &
    \subfigure{ 
    \resizebox{0.21\textwidth}{!}{%
    \begin{tikzpicture}
    \begin{axis}
    [
    ymode=log,
    ymajorgrids=true,
    ybar,
    /pgf/bar width=3.5pt,
    legend style={at={(0.5,-0.3)}, 
    anchor=south,legend columns=-1},   
    symbolic x coords={{1000},{2000},{4000},{6000},{8000},{10000}},
    xtick=data,
    ]
        \addplot  table[x=D,y=HG_local] {tables/timecost2.csv}; \addlegendentry{\small Local}
        \addplot table[x=D,y=HG_global] {tables/timecost2.csv}; \addlegendentry{\small Global}
        \addplot  table[x=D,y=HG_total] {tables/timecost2.csv}; \addlegendentry{\small $HG_+$}
        \legend{Local, Global, $HG_+$}
    \end{axis}
    \end{tikzpicture}
    }
    }
    \end{tabular}
    \caption{Efficiency comparison - time costs (s) ($\epsilon_G=\epsilon_L=0.5$)}
    \label{fig:timecost}
\end{figure}
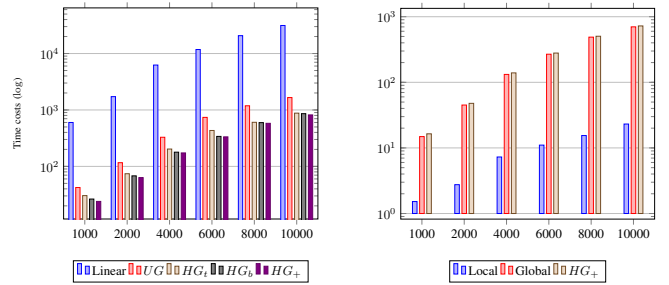

In this section, we evaluate the effectiveness of our proposed hierarchical grid index ($HG$) with different kNN strategies compared to the basic linear scan approach and the single-level uniform grid. We vary the dataset size (i.e., total number of objects/trajectories) within \{1000, 2000, 4000, 6000, 8000, 10000\}, and report the overall efficiency of these search algorithms in \Cref{fig:timecost}. 
It can be seen that our hierarchical grid index $HG$ is quite powerful in filtering unpromising candidates during $K$-nearest neighbor search, leading to dramatically higher efficiency than the \textit{Linear} approach (reduce the time cost by more than $30$ times). Compared to the single-level uniform grid index ($UG$), hierarchical structure helps a lot to properly organize the trajectory segments with variable lengths as well as discard misleading information, contributing to two times improvement in efficiency. Regarding the three search strategies, namely top-down ($HG_t$), bottom-up ($HG_b$) and bottom-up-down ($HG_+$), $HG_b$ is slightly faster than $HG_t$, both of which are outperformed by $HG_+$. Starting from the finest-grained grid cell containing the query point can accelerate the process of threshold shrinking, which in turn enhances the pruning power of $HG_+$ and avoids more unnecessary calculations.

We also compare the time cost of local intra- and global inter-trajectory modification. According to \Cref{fig:timecost}, global alteration dominates the overall process, occupying $90\%+$ of the total time approximately. As analyzed in \Cref{subsec:index}, global and location modifications conduct kNN search with the $HG$ index for $O(dn)$ and $O(mn)$ times respectively, where $n$ represents the dataset size, $m$ the reserved signature size, and $d$ the dimensionality of signature set $\mathcal{P}$ generated by grouping the signature points of each trajectory in the entire dataset, and thus $d=m\times n$ in the worst case, leading to $O(mn^2)$ search in the hierarchical grid index. As a result, global alteration is much more time-consuming than the local one. In future, we will further improve the efficiency of global alteration by early pruning unpromising trajectories based on their bounding box.

\section{Conclusion}
In this work, we propose a frequency-based randomization model with the differential privacy guarantee to protect individual privacy that might be exposed from spatial trajectories. Specifically, two independent perturbation mechanisms are provided by injecting non-trivial Laplace noises into the local point frequency and global trajectory frequency distributions over the private signature points, respectively. Moreover, the task of trajectory modification based on the distorted frequency distributions with minimum utility loss is formalized as $K$-nearest trajectory (segment) search problems. In order to support efficient search, we leverage the locality property of trajectories and design a hierarchical grid indexing structure with a novel bottom-up-down search algorithm, obtaining superior performance in a large-scale trajectory dataset. Empirically, our anonymization model outperforms both DP-related methods and signature-based approaches developed in the literature, achieving a satisfactory balance between privacy protection and utility preservation, and the frequency-perturbed trajectories can hardly be recovered by map-matching techniques. 

\label{sec:conclude}

\section*{Acknowledgment}
This work was partially supported by the Australian Research Council (DP200103650 and LP180100018) and the Natural Science Foundation of China (Grant No. 62072125).

\balance

\bibliographystyle{IEEEtran}
\bibliography{reference}

\end{document}